\documentclass[%
 aip,
 jmp,
preprint,%
]{revtex4-1}

\usepackage[utf8]{inputenc}
\usepackage[T1]{fontenc}

\usepackage{graphicx}%
\usepackage{dcolumn}%
\usepackage{bm}%

\usepackage[utf8]{inputenc}
\usepackage[T1]{fontenc}
\usepackage{mathptmx}
\usepackage{amsmath}
\usepackage{amsthm}
\usepackage{amsfonts}
\usepackage{mathrsfs}
\usepackage{amssymb} 
\usepackage{bbm,dsfont} 
\usepackage{units}
\usepackage{stmaryrd}   
\usepackage{verbatim}
\usepackage{enumerate} 
\usepackage{wasysym}
\usepackage{tikz}
\usetikzlibrary{calc}
\usepackage{tabularx}
\usepackage{accents}

\usepackage{calrsfs}

\usepackage{xcolor}
\usepackage{hyperref} %
\usepackage%
{hyperref} %
\hypersetup{
    colorlinks,
    linkcolor={blue!80!black},%
    citecolor={green!30!black},
    urlcolor={blue!80!black}
}

\newcommand{\bieee}{\begin{IEEEeqnarray}{rCl}}
\newcommand{\eieee}{\end{IEEEeqnarray}}
\newcommand{\prob}[1]{\Pr\left(#1\right)}
\newcommand{\given}{\mid}
\newcommand{\cprob}[2]{\Pr\left(#1\given #2\right)}

\renewcommand{\mathbbm}[1]{\text{\usefont{U}{bbm}{m}{n}#1}} 
\newcommand{\eps}{\varepsilon}

\newcommand{\norm}[1]{\left\lVert#1\right\rVert}
\newcommand{\trace}{\mathrm{Tr}}

\newcommand{\identity}{\mathbbm{1}}
\newcommand{\kb}[1]{ | #1 \rangle\langle #1 | } 

\newcommand{\ie}{\emph{i.e.} }
\newcommand{\eg}{\emph{e.g.} }
\newcommand{\etal}{\emph{et al.} }
\newcommand{\cf}{\emph{cf.} }

\newcommand{\tm}{\widetilde{m}}	
\newcommand{\tM}{\widetilde{M}}

\newcommand{\htm}{\hat{m}}

\newcommand{\hM}{\widehat{M}}

\newcommand{\Aset}{\mathcal{A}}

\newcommand{\Dset}{\mathcal{D}}
\newcommand{\Fset}{\mathcal{F}}

\newcommand{\Hset}{\mathcal{H}}

\newcommand{\Mset}{\mathcal{M}}
\newcommand{\Pset}{\mathcal{P}}
\newcommand{\Uset}{\mathcal{U}}

\newcommand{\Xset}{\mathcal{X}}
\newcommand{\Yset}{\mathcal{Y}}
\newcommand{\Zset}{\mathcal{Z}}
\newcommand{\Eset}{\mathcal{E}}

\theoremstyle{remark}	\newtheorem{theorem}{Theorem}
\theoremstyle{remark}	\newtheorem{lemma}[theorem]{Lemma}
\theoremstyle{remark}	
\theoremstyle{remark}	
\theoremstyle{remark} \newtheorem{definition}{Definition}
\theoremstyle{remark} \newtheorem{remark}{Remark}
\theoremstyle{remark}

\newcommand{\tset}{\Aset^{\delta}}													

\newcommand{\channel}{\mathcal{N}}

\newcommand{\inC}{\mathsf{C}}

\newcommand{\opR}{\mathbb{R}}

\begin{document}

\title[Quantum Broadcast Channels with Cooperating Decoders]{Quantum Broadcast Channels with Cooperating Decoders: An Information-Theoretic Perspective on Quantum Repeaters}

\author{Uzi Pereg}
\email{uzi.pereg@tum.de}
\author{Christian Deppe}%
\email{christian.deppe@tum.de}
\affiliation{Institute of Communication Engineering, Technical University of Munich.}

\author{Holger Boche}
\email{boche@tum.de}
\affiliation{Theoretical Information Technology, Technical University of Munich.}
\affiliation{Munich Center for Quantum Science and Technology (MCQST).} %
\homepage[Part of this work has been presented  at the IEEE International Symposium on Information Theory (ISIT 2021),\\
 Melbourne, Australia, July 12 – 20, 2021.]{}

\date{\today}

\begin{abstract}
Communication over a quantum broadcast channel with cooperation between the receivers is considered.
The first form of cooperation addressed is classical conferencing, where Receiver 1 can send classical messages to Receiver 2.
Another cooperation setting involves quantum conferencing, where Receiver 1 can teleport a quantum state to Receiver 2.
When Receiver 1 is not required to recover information and its sole purpose is to help the transmission to Receiver 2, the model reduces to the quantum primitive relay channel.
The quantum conferencing setting is intimately related to quantum repeaters, as the sender, Receiver 1, and Receiver 2 can be viewed as the transmitter, the repeater, and the destination receiver, respectively.
We develop lower and upper bounds on the capacity region in each setting. In particular,
the cutset upper bound and the decode-forward lower bound are derived  for the primitive relay channel. %
Furthermore, we present an entanglement-formation lower bound,  
where a virtual channel is simulated through the conference link.
At last, we show that as opposed to the multiple access channel with entangled encoders,  entanglement between decoders does not increase the classical communication rates for the broadcast dual.
\end{abstract}


\maketitle

\section{Introduction}
Attenuation in optical fibers poses a great challenge for long-distance quantum communication protocols, including both current applications such as quantum key distribution \cite{JTNMVL:09p}, as well as future implementation of the quantum internet
\cite{BSDP:19c} and quantum networks in general \cite{BFSDBFJ:20b}.
Quantum repeaters have been proposed as a potential solution where the distance is divided into smaller segments with quantum repeaters at the intermediate stations \cite{BriegelDurCiracZoller:98p}. %
In its simplest form, the process begins with using quantum communication and entanglement distillation to prepare two pairs of qubits at maximally entangled states, namely, $|\Phi_{AP_1}\rangle$ between the sender and the repeater, and $|\Phi_{P_2 B}\rangle$ between the repeater and the receiver. %
At the next stage, the repeater teleports the quantum state of $P_1$ onto $B$ thus swapping the entanglement such that $A$ and $B$ are now entangled at a distance twice that of the initial entangled pairs.
Experimental implementation of the elementary building blocks for quantum repeaters has recently been considered by
van Loock \etal \cite{vanLoockAltBecherBensonBocheDeppe:20p} in platforms based on quantum dots \cite{GiesFlorianSteinhoffJahnke:17b}, trapped ions \cite{KMSKHL:19p}, and color centers in diamond \cite{NSBMLKSCBR:19p,RYGRHHWE:19p}.  
Here, we will give an information-theoretic perspective that can be associated with such a network.

The cross-disciplinary  field of quantum information processing and communication is rapidly evolving in both practice and theory 
\cite{DowlingMilburn:03p,ZBBBBCCDEE:05p,JKLGD:13p,%
BecerraFanMigdall:15p,YCLZRC:17p,ZDSZSG:17p,LYZGCZHLJL:19p,
PERLHPCVV:20p}.
Quantum information theory is the natural extension of the classical  theory. Nevertheless, 
this generalization reveals astonishing phenomena with no parallel in classical communication \cite{GyongyosiImreNguyen:18p}. For example, 
pairing two memoryless quantum channels, each with zero quantum capacity, can result in a nonzero quantum capacity
\cite{SmithYard:08p}. This property is referred to as super-activation.
It should be noted that super-activation has also been demonstrated in recent years for classical channels in advanced settings, such as %
secure message transmission over a wiretap channel with a jammer \cite{BocheShaefer:13p}
and identification over a discrete memoryless channel with feedback \cite{BocheSchaeferPoor:20p}.
Nevertheless, super-activation does not %
occur in the fundamental model of a %
classical one-way memoryless channel.

Communication over quantum channels can be separated into different tasks and categories.
 For classical information transmission, a regularized (``multi-letter")  formula for the capacity of a quantum channel without assistance was established by Holevo \cite{Holevo:98p}, and Schumacher \etal \cite{SchumacherWestmoreland:97p}. %
Although the calculation of such a formula is intractable in general, it provides computable lower bounds, and there are special cases where the capacity can be computed exactly. 
The reason for this difficulty is that the Holevo information is not necessarily additive \cite{Holevo:12b}. %
A similar difficulty occurs with the transmission of quantum information. %
A regularized formula for the quantum capacity is given in Refs. \onlinecite{BarnumNielsenSchumacher:98p,Loyd:97p,Shor:02l,Devetak:05p}. %
A computable %
formula is obtained in the special case where the channel is  degradable \cite{DevetakShor:05p}, or belongs to the more general class of less noisy channels \cite{Watanabe:12p}. %
 Quantum communication can also be used for the purpose of entanglement generation \cite{Devetak:05p,BjelakovicBocheNotzel:09p,WildeHsieh:10c}.

Another scenario of interest is when the transmitters and receivers %
are provided with entanglement resources a priori \cite{NielsenChuang:02b,BocheJanssenKaltenstadler:17p,PeregDeppeBoche:21p}.
While entanglement can be used to produce shared randomness, it is a much more powerful aid \cite{Wilde:17b,ChitambarGour:19p,BFSDBFJ:20b}. 
In particular, super-dense coding \cite{BennetWiesner:92p} is a well-known communication protocol where two classical bits are transmitted using a single use of a noiseless qubit channel and a maximally entangled pair that is shared between the transmitter and the receiver. Thereby, transmitter-receiver entanglement assistance doubles the transmission rate of classical messages over a noiseless qubit channel. 
 The entanglement-assisted capacity of %
a noisy quantum channel was fully characterized by Bennet \etal 
\cite{BennettShorSmolin:99p,BennettShorSmolin:02p} in terms of the quantum mutual information. 
In the other direction, i.e. using information measures to understand quantum physics, the quantum mutual information plays a role in investigating the entanglement structure of quantum field theories %
\cite{Swingle:10a,PanJing:08p,CasiniHuertaMyersYale:15p,AgonGaulkner:16p}.

There are communication settings where entanglement resources can even increase the capacity of a \emph{classical} channel. %
In particular, Leditzky et al. \cite{LeditzkyAlhejjiLevinSmith:20p} have recently 
 shown that entanglement between two transmitters can strictly increase the achievable rates for a classical multiple access channel. The
channel construction in Ref. \onlinecite{LeditzkyAlhejjiLevinSmith:20p} is based on  a pseudo-telepathy game \cite{BrassardBroadbentTapp:05p} where quantum strategies guarantee a certain win and outperform classical strategies,   extending ideas by N\"otzel \cite{Notzel:19a} and Quek and Shor
\cite{QuekShor:17p}.
Entanglement assistance  
has striking effects in different communication  games and their security applications as well \cite{CHSH:69p,PappaChaillouxWehnerDiamantiKerenidis:12p,VaziraniVidick:14p,JaiWeiWuGuo:20a,JiNatarajanVidickWrightYuen:20a,LeditzkyAlhejjiLevinSmith:20p}.
Furthermore, entanglement can assist in %
the transmission of quantum information. Given a classical channel with transmitter-receiver entanglement resources,
 qubits can be sent %
at half the rate of classical bits %
by employing the teleportation protocol \cite{BennettBrassardJozsaPeres:93p}. %

Quantum broadcast and multiple access channels were studied in various settings, as \eg in
 Refs. \onlinecite{YardHaydenDevetak:11p,SavovWilde:15p,
RadhakrishnanSenWarsi:16p,WangDasWilde:17p,DupuisHaydenLi:10p,Dupuis:10z,HircheMorgan:15c,SeshadreesanTakeokaWilde:16p,BaumlAzuma:17p,HeinosaariMiyadera:17p,BochCaiDeppe:15p,Hirche:12z,XieWangDuan:18c,Palma:19p,AnshuJainWarsi:19p1,ChengDattaRouze:19a}
and
\onlinecite{Winter:01p,Klimovitch:01c,Yard:05z,HsiehDevetakWinter:08p,YardHaydenDevetak:08p,CzekajHorodecki:09p,BocheNoetzel:14p,DiadamoBoche:19a}. 
Yard \etal \cite{YardHaydenDevetak:11p} derived the superposition inner bound and determined the capacity region for the degraded classical-quantum  broadcast channel.
By the monogamy property of quantum entanglement \cite{KoashiWinter:04p}, the sender's system cannot be in a maximally entangled state with both receivers simultaneously. 
However, different forms of entanglement can be generated. In particular, Yard \etal \cite{YardHaydenDevetak:11p} characterize the entanglement-generation rates for  GHZ states.
Wang \etal \cite{WangDasWilde:17p} used the previous characterization to determine the capacity region for Hadamard broadcast channels.
Dupuis \etal \cite{DupuisHaydenLi:10p,Dupuis:10z} developed the entanglement-assisted version of Marton's region for users with independent messages.
Bosonic broadcast channels are considered in Refs. \onlinecite{GuhaShapiro:07c,GuhaShapiroErkmen:07p,DePalmaMariGiovannetti:14p,TakeokaSeshadreesanWilde:16c,TakeokaSeshadreesanWilde:17p}. 
The quantum broadcast and multiple access channels with confidential messages were recently considered in Refs.  \onlinecite{SalekHsiehFonollosa:19a}-\onlinecite{SalekHsiehFonollosa:19c} and  \onlinecite{AghaeeAkhbari:19c}-\onlinecite{BochJanssenSaeedianaeeni:20p}, respectively. %
An equivalent description of the super-activation phenomenon \cite{SmithYard:08p} is that there exists a broadcast channel such that 
the sum-rate capacity with full cooperation between the receivers is positive while the capacities of the marginal channels are both zero.

Savov \etal \cite{SavovWildeVu:12c,Savov:12z} derived a partial decode-forward lower bound for the (non-primitive) classical-quantum relay channel, where the relay encodes information in a strictly-causal manner. 
Recently, Ding \etal \cite{DingGharibyanHaydenWalter:20p} generalized those results and established the cutset, multihop, and coherent multihop bounds for the classical-quantum relay channel. %
Communication with the help of environment measurement can be modelled by a quantum channel with a classical relay in the environment \cite{HaydenKing:04a}.
Considering this setting, Smolin \etal \cite{Smolin:05p} and Winter \cite{Winter:05a} determined the environment-assisted quantum capacity
and classical capacity, %
respectively. %
Savov \etal \cite{SavovWildeVu:12c} further discussed future research directions of interest (see Sec. V in Ref. 
\onlinecite{SavovWildeVu:12c}), and pointed out that quantum communication scenarios over the relay channel may have applications for the design of quantum repeaters (see also Ref. \onlinecite{DingGharibyanHaydenWalter:20p}).  Our aim is to fulfill this prevision.

In this paper, we consider quantum broadcast channels in different settings of cooperation between the decoders. Using those settings, we provide an information-theoretic framework for quantum repeaters. The first form of cooperation that we consider is classical conferencing, where Receiver 1 can send classical messages to Receiver 2. 
This can be viewed as the quantum version of the classical setting by Dabora and Servetto \cite{DaboraServetto:06p} (see also Ref. \cite{Steinberg:15c}). 
We provide a regularized characterization for the classical capacity region of the quantum broadcast channel with classical conferencing, and a single-letter formula for Hadamard broadcast channels \cite{WangDasWilde:17p}.
Next, we consider quantum conferencing, where Receiver 1 can teleport a quantum state to Receiver 2.
We develop inner and outer bounds on the quantum capacity region with quantum conferencing, characterizing the tradeoff between the communication rates $Q_1$ and $Q_2$ to Receiver 1 and Reciever 2, respectively, as well as the conferencing capacity $\inC_{Q,12}$.
The case where Receiver 1 is not required to recover information and its sole purpose is to help the transmission to Receiver 2,   reduces to the model of the primitive relay channel \cite{Kim:07c}, for which the decode-forward lower bound and cutset upper bound follow as a consequence. In addition, we establish an entanglement-formation lower bound,  
where a virtual channel is simulated through the conference link, following the results of Berta \etal \cite{BertaBradaoChristandlWehner:13p} on quantum channel simulation.

  The quantum conferencing setting is intimately related to quantum repeaters, as the sender, Receiver 1, and Receiver 2 can be viewed as the transmitter, the repeater, and the destination receiver, respectively, in the  repeater model. In particular, the sender can employ quantum communication to Receiver 1 (the repeater)  in order to prepare a maximally entangled pair $|\Phi_{A P_1}\rangle$, which consists of $nQ_1$ entangled bits (ebits).
Given entanglement between the receivers, we also have a maximally entangled pair $|\Phi_{P_2 B}\rangle$, which consists of $n\inC_{Q,12}$ ebits, shared between the repeater and the destination receiver.  
 Then, the repeater can swap his entanglement by using the classical conferencing link to
 teleport the state of $P_1$ onto $B$ thus swapping the entanglement such that $A$ and $B$ are now entangled. Hence our results provide an information-theoretic analysis characterizing the achievable rates of ebits that can be generated in each stage.
As our model includes direct transmission from $A$ to $B$ as well, our results exhibit the tradeoff between repeaterless communication and  communication via the repeater. Other relay channel models for quantum repeaters can also be found in 
Refs. \onlinecite{GyongyosiImre:12c,JRXYLM:12p,GyongyosiImre:14c,Pirandola:16a,GhalaiiPirandola:20a}. %

At last, we compare between entanglement cooperation for the multiple access channel and the broadcast channel.
The duality between the multiple access channel and the broadcast channel has emerged as a prominent tool in the study of wireless communication systems \cite{JindalVishwanathGoldsmith:04p,VishwanathTse:03p,WeingartenSteinbergShamai:06p1}.
We show that as opposed to the multiple access channel with entangled transmitters \cite{LeditzkyAlhejjiLevinSmith:20p},  entanglement between the receivers cannot enlarge the classical capacity region
of a broadcast channel. Furthermore, this property extends to any pair of non-signaling correlated resources that are shared between the receivers.
 Consequently, the broadcast dual to the multiple access channel property by Leditzky et al. \cite{LeditzkyAlhejjiLevinSmith:20p} does not hold. %
Hence, our result reveals a fundamental asymmetry and demonstrates the limitations of the duality between the broadcast channel and the multiple access channel.

The paper is organized as follows. 
In Sec.~\ref{sec:def}, we begin with the basic definitions.  %
In Sec.~\ref{sec:Coding}, we present three coding scenarios for the quantum broadcast channel with cooperation between the receivers. In particular, we consider classical communication when the receivers share entanglement resources a priori
(Subsec.~\ref{subsec:BcodingClE}); classical communication over the quantum broadcast channel with a classical conference link from Receiver 1 to Receiver 2 (Subsec.~\ref{subsec:BcodingCf}); and quantum communication when Receiver 1 can teleport a quantum state to Receiver 2 via conferencing (Subsec.~\ref{subsec:BcodingQF}). The quantum primitive relay channel is presented as a special case as well.  Our results for classical conferencing are given in Sec.~\ref{sec:Clf}.
Next, our main results on quantum conferencing are derived in Sec.~\ref{sec:Qf}, for the quantum broadcast channel (\ref{subsec:Qfachieve},\ref{subsec:outB}), and the quantum primitive relay channel (\ref{subsec:qPrl}).
Sec.~\ref{sec:Qf} is concluded with the resulting observations on the quantum repeater.
In Sec.~\ref{sec:EdecCl}, we show that the broadcast dual to the multiple access channel property by Leditzky et al. \cite{LeditzkyAlhejjiLevinSmith:20p} does not holds, as entanglement between receivers cannot enlarge the classical capacity region.
We conclude with a summary and discussion in Sec.~\ref{sec:summary}. %

\section{Definitions}%
\label{sec:def}
\subsection{Notation, States, and Information Measures}
\label{subsec:notation}
 We use the following notation conventions. %
Script letters $\Xset,\Yset,\Zset,...$ are used for finite sets.
Lowercase letters $x,y,z,\ldots$  represent constants and values of classical random variables, and uppercase letters $X,Y,Z,\ldots$ represent classical random variables.  
 The distribution of a  random variable $X$ is specified by a probability mass function (pmf) 
	$p_X(x)$ over a finite set $\Xset$. %
 We use $x^j=(x_1,x_{2},\ldots,x_j)$ to denote  a sequence of letters from $\Xset$. %
 A random sequence $X^n$ and its distribution $p_{X^n}(x^n)$ are defined accordingly. 

The state of a quantum system $A$ is a density operator $\rho$ on the Hilbert space $\Hset_A$.
A density operator is an Hermitian, positive semidefinite operator, with unit trace, \ie 
 $\rho^\dagger=\rho$, $\rho\succeq 0$, and $\trace(\rho)=1$.
The state is said to be pure if $\rho=\kb{\psi}$, for some vector $|\psi\rangle\in\Hset_A$, where
$\langle \psi |$ %
is the Hermitian conjugate of $|\psi\rangle$. 
In general, a density operator has a spectral decomposition, %
\begin{align}
\rho=\sum_{x\in\Xset} p_X(x) \kb{ \psi_x } %
\end{align}
where $\Xset=\{1,2,\ldots,|\Hset_A|\}$, $p_X(x)$ is a probability distribution over $\Xset$, and $\{ |\psi_x\rangle \}_{x\in\Xset}$ forms an orthonormal basis of the Hilbert space $\Hset_A$.
A measurement of a quantum system is any set of operators $\{\Lambda_j \}$ that forms a positive operator-valued measure (POVM), \ie
the operators are positive semi-definite and %
$\sum_j \Lambda_j=\identity$, where $\identity$ is the identity operator \cite{NielsenChuang:02b}.
According to the Born rule, if the system is in state $\rho$, then the probability of the measurement outcome $j$ is given by $p_A(j)=\trace(\Lambda_j \rho)$.
The trace distance between two density operators $\rho$ and $\sigma$ is $\norm{\rho-\sigma}_1$ where $\norm{F}_1=\trace(\sqrt{F^\dagger F})$.

Define the quantum entropy of the density operator $\rho$ as
$%
H(\rho) \triangleq -\trace[ \rho\log(\rho) ]
$, %
which is the same as the Shannon entropy %
associated with the eigenvalues of $\rho$.
Consider the state of a pair of systems $A$ and $B$ on the tensor product $\Hset_A\otimes \Hset_B$ of the corresponding Hilbert spaces.
Given a bipartite state $\sigma_{AB}$, %
define the quantum mutual information as
\begin{align}
I(A;B)_\sigma=H(\sigma_A)+H(\sigma_B)-H(\sigma_{AB}) \,. %
\end{align} 
Furthermore, conditional quantum entropy and mutual information are defined by
$H(A|B)_{\sigma}=H(\sigma_{AB})-H(\sigma_B)$ and
$I(A;B|C)_{\sigma}=H(A|C)_\sigma+H(B|C)_\sigma-H(A,B|C)_\sigma$, respectively.
The coherent information is then defined as
\begin{align}
I(A\rangle B)_\sigma=-H(A|B)_\sigma \,. %
\end{align}

A pure bipartite state %
is called \emph{entangled} if it cannot be expressed as the tensor product %
of two states %
in $\Hset_A$ and $\Hset_B$. %
The maximally entangled state %
between two systems %
of dimension $D$ %
is defined by
$%
| \Phi_{AB} \rangle = \frac{1}{\sqrt{D}} \sum_{j=0}^{D-1} |j\rangle_A\otimes |j\rangle_B %
$, where $\{ |j\rangle_A \}_{j=0}^{D-1}$ and $\{ |j\rangle_B \}_{j=0}^{D-1}$  %
are respective orthonormal bases. %
Note that $I(A;B)_{\kb{\Phi}}=2\cdot \log(D)$ and $I(A\rangle B)_{\kb{\Phi}}= \log(D)$.

The entanglement of formation of a joint state $\rho_{AB}$ is defined as \cite{BennettDiBincenzoSmolinWooters:96p,BertaBradaoChristandlWehner:13p}
\begin{align}
E_F(\rho_{AB})\triangleq \inf_{p_X(x) \,,\; |\psi^x_{AB}\rangle} H(A|X)_\rho
\end{align} 
where the infimum is over all pure state decompositions $\rho_{AB}=\sum_x p_X(x)\kb{\psi_{AB}^x}$.

\subsection{Quantum Broadcast Channel}
\label{subsec:Qchannel}
A quantum broadcast channel maps a quantum state at the sender system to a quantum state at the receiver systems. 
Here, we consider a channel with two receivers.
Formally, a quantum broadcast channel  is a   linear, completely positive, trace-preserving map 
$%
\channel_{ A\rightarrow B_1 B_2}  %
$ %
corresponding to a quantum physical evolution.
We assume that the channel is memoryless. That is, if the systems $A^n=(A_1,\ldots,A_n)$ are sent through $n$ channel uses, then the input state $\rho_{ A^n}$ undergoes the tensor product mapping
$%
\channel_{ A^n\rightarrow B_1^n B_2^n}\equiv  \channel_{ A\rightarrow B_1 B_2}^{\otimes n} %
$. %
The marginal channel $\channel^{(1)}_{A\rightarrow B_1}$ is defined by
\begin{align}
\channel_{A\rightarrow B_1}^{(1)}(\rho_A)=\trace_{B_2} \left( \channel_{ A\rightarrow B_1 B_2}(\rho_{A}) \right) 
\end{align}
for Receiver 1, and similarly $\channel_{A\rightarrow B_2}^{(2)}$ for Receiver 2. 
One may say that $\channel_{A\rightarrow B_1 B_2}$ is an extension of $\channel^{(1)}_{A\rightarrow B_1}$ and $\channel^{(2)}_{A\rightarrow B_2}$.
We will consider a broadcast channel with conferencing where Receiver 1 can transmit classical information to Receiver 2 using a noiseless communication link of capacity $\inC_{12}$. We will denote this classical communication channel by 
$CC_{G\rightarrow G'}$, where $G$ and $G'$ represent the registers that store the conference message transmitted from Receiver 1 and received at 
Receiver 2, respectively. The transmitter, Receiver 1, and Reciever 2 are often called Alice, Bob 1, and Bob 2.

A quantum broadcast channel has a Kraus representation,
\begin{align}
\channel_{ A\rightarrow B_1 B_2}(\rho_{A})=\sum_j N_j \rho_{A} N_j^\dagger 
\end{align}
for some set of operators $N_j$ such that $\sum_j N_j^\dagger N_j=\identity$. %

\begin{remark}
The classical broadcast channel is the special case where the input and the outputs can be represented by classical random variables
$X$ and $Y_1,Y_2$, respectively, while  the Kraus operators are
$N_{x,y_1,y_2}=\sqrt{P_{Y_1 Y_2|X}(y_1,y_2|x)} |y_1,y_2\rangle \langle x|$ for some
probability kernel $P_{Y_1 Y_2|X}$ and orthonormal bases $\{|x\rangle\}$,$\{|y_1,y_2\rangle%
\}$. Therefore,
given an input $x\in\Xset$, the output state of a classical broadcast channel is
\begin{align} 
\channel^{\,\text{Cl}}_{X\rightarrow Y_1 Y_2}(\kb{x})=\sum_{(y_1,y_2)\in\Yset_1\times \Yset_2} P_{Y_1 Y_2|X}(y_1,y_2|x) \kb{y_1,y_2} \,.
\end{align}
\end{remark}

\subsection{Degraded Broadcast Channel, Hadamard Broadcast Channel, and Degradable Marginals}
\label{subsec:LessNHadamard}
We will also be interested in the following special cases. %
\begin{definition}[Degraded broadcast channel and Hadamard broadcast channel {\cite{WangDasWilde:17p}}]
  A quantum broadcast channel $\channel_{A\rightarrow B_1 B_2}$ is called \emph{degraded} if there exists a degrading channel $\Pset_{B_1\rightarrow B_2}$ such that the marginals satisfy the following relation,
	\begin{align}
	\channel^{(2)}_{A\rightarrow B_2}=\Pset_{B_1\rightarrow B_2}\circ\channel^{(1)}_{A\rightarrow B_1} \,.
	\end{align}
In this case, we say that $\channel^{(2)}$ is degraded with respect to $\channel^{(1)}$.
A quantum-classical-quantum degraded channel $\channel^{\,\text{H}}_{A\rightarrow Y_1 B_2}$ is called a Hadamard broadcast channel.  
\end{definition}
Intuitively, if a broadcast channel is degraded, then  the output state of Receiver 2 is a noisy version of that of Receiver 1. 
A Hadamard broadcast channel $\channel^{\,\text{H}}_{A\rightarrow Y_1 B_2}$ can be viewed as a measure-and-prepare channel where the marginal channel $\channel^{(1)}_{A\rightarrow Y_1}$ acts as a measurement device, while the degrading channel $\Pset_{Y_1\rightarrow B_2}$ corresponds to state preparation.  In this case, the marginal quantum channel $\channel^{(2)}_{A\rightarrow B_2}$ of Receiver 2 is said to be entanglement-breaking \cite{Shor:02p}.

Next, we define a broadcast channel with degradable marginals.
Every point-to-point quantum channel $\Mset_{A\rightarrow B}$ has an isometric extension $\Uset^{\Mset}_{A\rightarrow BE}(\rho_{A})=U\rho_{A} U^\dagger$, also called  a Stinespring dilation, 
where the operator $U$ is an isometry, \ie $ U^\dagger U=\identity$ (see Ref. \onlinecite[Sec. VII]{BocheCaiCaiDeppe:14p}). %
The system $E$ is often associated with the decoder's environment, or with a malicious eavesdropper in  the wiretap channel model  \cite{Devetak:05p}.
The channel $\widehat{\Mset}_{A\rightarrow E}(\rho_{A})=\trace_B( U\rho_{A} U^\dagger )$ is called the complementary channel for
${\Mset}_{A\rightarrow B}$.
\begin{definition}[Degradable marginals]
 A point-to-point quantum channel $\Mset_{A\rightarrow B}$ is called \emph{degradable} if there exists an isometric extension such that the complementary channel $\widehat{\Mset}_{A\rightarrow E}$ is degraded with respect to $\Mset_{A\rightarrow B}$. In other words, the channel to the environment is degraded with respect to the channel to the receiver.
We say that the quantum broadcast channel $\channel_{A\rightarrow B_1 B_2}$ has degradable marginals if both marginals 
$\channel^{(1)}_{A\rightarrow B_1}$ and $\channel^{(2)}_{A\rightarrow B_2}$ are degradable.
\end{definition}
Examples of degradable quantum channels include the erasure channel and the dephasing channel \cite{DevetakShor:05p}.

\section{Coding for The Broadcast Channel}
\label{sec:Coding}
We consider different broadcast scenarios with cooperation between the decoders, where the transmitted information can be  classical or quantum, with entanglement resources or without, and when conferencing between the receivers is available or not.
\begin{center}
\begin{figure}[ht!]
\includegraphics[scale=0.75,trim={2cm 9cm -1cm 9cm},clip]{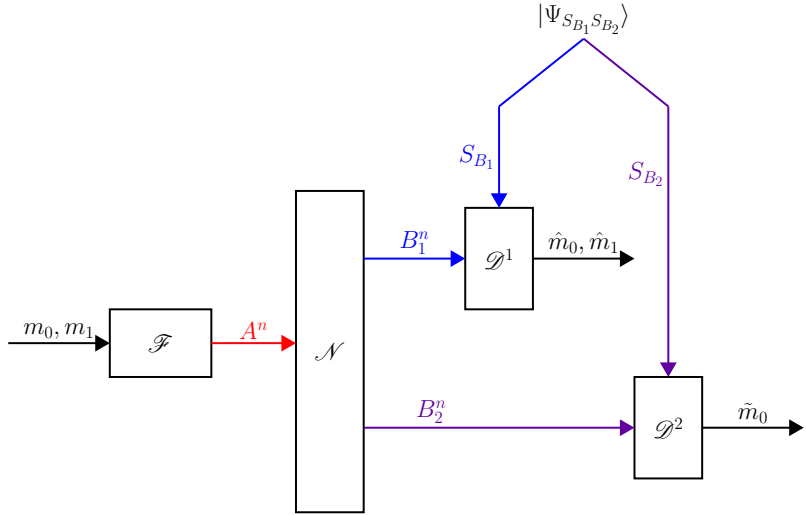} %
\caption{Classical coding for a quantum broadcast channel $\channel_{A\rightarrow B_1 B_2}$ with shared entanglement between the decoders and degraded message sets. The quantum systems of Alice, Bob 1, and Bob 2 are marked in red, blue, and purple, respectively. 
Bob 1 and Bob 2 share entanglement resources in the systems $S_{B_1}$ and $S_{B_2}$, respectively.
Alice encodes the messages $m_0$ and $m_1$ by applying the encoding map $\Fset_{  M_0 M_1 \rightarrow A^n}$ to the respective registers $M_0$ and $M_1$ which store the messages. 
Then, she transmits the systems $A^n$ over the broadcast channel. %
The decoder $\Dset^1$ of Bob $1$ receives the channel output systems $B_1^n$, and estimates the common and private messages by performing a decoding measurement on the systems $B_1^n$ and $S_{B_1}$, using a POVM $\{\Lambda^{m_0,m_1}_{S_{B_1} B_1^n}\}  $. Similarly,  the decoder $\Dset^2$ of Bob $2$ %
estimates the common message by measuring a POVM $\{\Gamma^{m_0}_{S_{B_2} B_2^n}\}  $ 
on $B_2^n$ and $S_{B_2}$.
}
\label{fig:EAconfCl}
\end{figure}

\end{center}

\subsection{Classical Coding with Entangled Decoders}
\label{subsec:BcodingClE}
First, we consider a broadcast channel where Receiver 1 and Receiver 2 share entanglement resources.
We denote their entangled systems %
by $S_{B_1}$ and $S_{B_2}$, respectively.

\begin{definition} %
\label{def:ClcapacityE}
A $(2^{nR_0},2^{nR_1},%
n)$ classical  code for the quantum broadcast channel $\channel_{A\rightarrow B_1 B_2}$ with
degraded message sets and entangled decoders  consists of the following: 
\begin{itemize}
\item 
Two index sets  $[1:2^{nR_0}]$ and $ [1:2^{nR_1}]$, corresponding to the common message for both users and the private message of User 1, respectively;
\item
	an encoding map $\Fset_{M_0 M_1  \rightarrow A^n}$, where $M_0$ and $M_1$ are classical registers that store the common and
   private messages, %
	respectively;
\item
a pure entangled state $\Psi_{S_{B_1},S_{B_2}}$
	
\item	
two decoding POVMs,   $\{\Lambda^{m_0,m_1}_{S_{B_1} B_1^n}\}  $ for Receiver 1 and $\{\Gamma^{m_0}_{S_{B_2} B_2^n}\}  $ for Receiver 2, where the measurement outcome $m_k$ is an index in 	$[1:2^{nR_k}]$, for $k=0,1$.
\end{itemize}
We denote the code by $(\Fset,\Psi,\Lambda,\Gamma)$.
\end{definition}

The communication scheme is depicted in Figure~\ref{fig:EAconfCl}.  
The sender Alice has the systems  $A^n$, and the receivers Bob 1 and Bob 2 have the systems $B_1^n,S_{B_1}$  and $B_2^n,S_{B_2}$,  respectively. Alice chooses a common message $m_0\in [1:2^{nR_0}]$ that is intended for both users and a private message $m_1\in [1:2^{nR_1}]$ for Bob 1, and stores them in the classical registers $M_0$ and $M_1$, respectively.
She encodes the messages by applying the encoding map $\Fset_{  M_0 M_1 \rightarrow A^n}$ %
which results in an input state
$%
\rho^{m_0,m_1}_{A^n}= \Fset_{ M_0 M_1 \rightarrow A^n}( m_0,m_1  ) %
$, %
 and transmits the systems $A^n$ over $n$ channel uses of $\channel_{A\rightarrow B_1 B_2}$. Hence, the output state is
\begin{align}
\rho^{m_0,m_1}_{ B_1^n B_2^n}=\channel^{\otimes n} (\rho^{m_0,m_1}_{A^n}) \,.
\end{align}
Bob 1 receives the channel output systems $B_1^n$, combines them with his entangled system $S_{B_1}$, and applies the POVM 
$\{\Lambda^{m_0,m_1}_{S_{B_1} B_1^n}\}  $. Bob 1 then obtains from the measurement outcome
 an estimate of the message pair $(\htm_0,\htm_1)\in [1:2^{nR_0}]\times [1:2^{nR_1}]$.
 Similarly, Bob 2 finds an estimate of the common message $\tm_0\in [1:2^{nR_0}]$ by performing a measurement using $\{\Gamma^{m_0}_{S_{B_2} B_2^n}\}  $ on the output systems $B_2^n$ and his entangled system $S_{B_2}$.  
The conditional probability of error of the code,  %
given that the message pair $(m_0,m_1)$ was sent, is given by 
\begin{align}
P_{e|m_0,m_1}^{(n)}(\Fset,\Psi,\Lambda,\Gamma)&= 
1-\trace[  (\Lambda^{m_0,m_1}_{S_{B_1} B_1^n}\otimes  \Lambda^{m_0}_{S_{B_1} B_2^n})  \rho^{m_0,m_1}_{ B_1^n B_2^n} ] \,.
\end{align}

A $(2^{nR_0},2^{nR_1},n,\eps)$ classical code satisfies 
$%
P_{e|m_0,m_1}^{(n)}(\Fset,\Psi,\Lambda,\Gamma)\leq\eps $ %
 for %
all $(m_0,m_1)\in [1:2^{nR_0}]\times [1:2^{nR_1}]$.  %
A rate pair $(R_0,R_1)$ is called achievable with entangled decoders  if for every $\eps>0$ and sufficiently large $n$, there exists a 
$(2^{nR_0},2^{nR_1},n,\eps)$ code. 
 The classical capacity region %
is defined as the set of achievable pairs $(R_0,R_1)$ with entangled decoders.

One may also consider the broadcast channel with independent messages, i.e. when the common message $m_0$ is replaced by a private message $m_2$ that is intended for Bob 2, in which case Bob 1 is not required to decode this message.
In general, the capacity region with independent messages can be larger than with degraded message sets.
\begin{center}
\begin{figure}[ht!]
\includegraphics[scale=0.75,trim={1cm 10.5cm 0 11cm},clip]{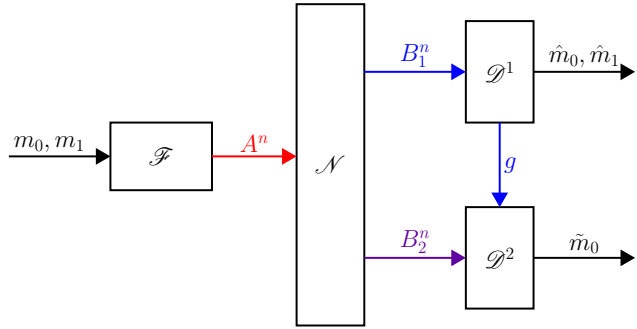} %
\caption{Classical coding for a quantum broadcast channel $\channel_{A\rightarrow B_1 B_2}$ with conferencing and degraded message sets. The quantum systems of Alice, Bob 1, and Bob 2 are marked in red, blue, and purple, respectively. 
There is a conferencing link between the decoders which allows Bob 1 to send a conferencing message to Bob 2 at a rate $\inC_{12}$.
Alice encodes the messages $m_0$ and $m_1$ by applying the encoding map $\Fset_{  M_0 M_1 \rightarrow A^n}$ to the respective registers $M_0$ and $M_1$ which store the messages. 
Then, she transmits the systems $A^n$ over the broadcast channel. %
The decoder $\Dset^1$ of Bob 1 receives the channel output systems $B_1^n$ and performs a measurement using the POVM $\{\Lambda^{m_0,m_1,g}_{B_1^n}\}  $. Bob 1 then obtains from the measurement outcome an estimate of the message pair  and a conference message.
 Next, Bob 1 sends the conference message $g$ to Bob 2. Given the conference message $g$, the decoder $\Dset^2$ of Bob 2 chooses a measurement POVM $\{\Gamma^{m_0}_{ B_2^n|g}\}  $ to perform on the channel output systems $B_2^n$, producing an estimate of the common message as the measurement outcome.
}
\label{fig:BconfCl}
\end{figure}

\end{center}

\subsection{Classical Coding with Conferencing}
\label{subsec:BcodingCf}
Another form of cooperation between the decoders involves conferencing. 
We consider a broadcast channel where 
Receiver 1 can transmit information to Receiver 2 using a classical conferencing link of capacity $\inC_{1 2}$.

\begin{definition} %
\label{def:Clcapacity}
A $(2^{nR_0},2^{nR_1},%
n)$ classical  code for the quantum broadcast channel $\channel_{A\rightarrow B_1 B_2}$ with
degraded message sets and conferencing  consists of the following: 
\begin{itemize}
\item 
Three index sets  $[1:2^{nR_0}]$, $ [1:2^{nR_1}]$, and $[1:2^{n\inC_{12}}]$, corresponding to the common message for both users, the private message of User 1, and the conference message, respectively;
\item
	an encoding map $\Fset_{M_0 M_1  \rightarrow A^n}$, where $M_0$ and $M_1$ are classical registers that store the common and
   private messages, %
	respectively;
	
\item	
	a decoding POVM   $\{\Lambda^{m_0,m_1,g}_{B_1^n}\}  $ for Receiver 1, where the measurement outcome is a triplet of indices $(m_0,m_1,g)$ in 	$[1:2^{nR_0}]\times [1:2^{nR_1}]\times [1:2^{\inC_{12}}]$; and
\item
 a collection of decoding POVMs $\{\Gamma^{m_0}_{ B_2^n|g}\}  $, $g\in[1:2^{\inC_{12}}]$, for Receiver 2, where the measurement outcome is an index in $[1:2^{nR_0}]$.
\end{itemize}
We denote the code by $(\Fset,\Lambda,\Gamma)$.
\end{definition}

The communication scheme is depicted in Figure~\ref{fig:BconfCl}.  
The sender Alice has the systems  $A^n$, and the receivers Bob 1 and Bob 2 have the systems $B_1^n$  and $B_2^n$,  respectively. Alice chooses a common message $m_0\in [1:2^{nR_0}]$ and a private message $m_1\in [1:2^{nR_1}]$ for Bob 1, and stores them in the classical registers $M_0$ and $M_1$, respectively.
She encodes the messages by applying the encoding map $\Fset_{  M_0 M_1 \rightarrow A^n}$ %
which results in an input state
\begin{align}
\rho^{m_0,m_1}_{A^n}= \Fset_{ M_0 M_1 \rightarrow A^n}( m_0,m_1  ) %
\end{align}
 and transmits the systems $A^n$ over %
$n$ channel uses of $\channel_{A\rightarrow B_1 B_2}$. Hence, the output state is
\begin{align}
\rho^{m_0,m_1}_{ B_1^n B_2^n}=\channel_{ A\rightarrow B_1 B_2}^{\otimes n} (\rho^{m_0,m_1}_{A^n}) \,.
\end{align}
Bob 1 receives the channel output systems $B_1^n$ and applies the POVM $\{\Lambda^{m_0,m_1,g}_{B_1^n}\}  $. Bob 1 then obtains from the measurement outcome
 an estimate of the message pair $(\htm_0,\htm_1)\in [1:2^{nR_0}]\times [1:2^{nR_1}]$ and a conference message $g\in [1:2^{n\inC_{12}}]$.
 Next, Bob 1 sends the conference message $g$ to Bob 2. Given the conference message $g$, Bob 2 chooses a POVM $\{\Gamma^{m_0}_{ B_2^n|g}\}  $ to perform on the channel output systems $B_2^n$, producing an estimate of the common message $\tm_0\in [1:2^{nR_0}]$ as the measurement outcome. 
The conditional probability of error of the code,  %
given that the message pair $(m_0,m_1)$ was sent, is given by 
\begin{align}
P_{e|m_0,m_1}^{(n)}(\Fset,\Lambda,\Gamma)&= 
1-\sum_{g=1}^{2^{n\inC_{12}}}\trace[  (\Lambda^{m_0,m_1,g}_{B_1^n}\otimes  \Lambda^{m_0}_{B_2^n|g})  \rho^{m_0,m_1}_{ B_1^n B_2^n} ] \,.
\end{align}

A $(2^{nR_0},2^{nR_1},n,\eps)$ classical code satisfies 
$%
P_{e|m_0,m_1}^{(n)}(\Fset,\Lambda,\Gamma)\leq\eps $ %
 for %
all $(m_0,m_1)\in [1:2^{nR_0}]\times [1:2^{nR_1}]$.  %
A rate pair $(R_0,R_1)$ is called achievable with conferencing  if for every $\eps>0$ and sufficiently large $n$, there exists a 
$(2^{nR_0},2^{nR_1},n,\eps)$ code. 
 The classical capacity region $\opR_{\text{Cl}}(\channel)$ %
is defined as the set of achievable pairs $(R_0,R_1)$ with conferencing.

\begin{remark}
The setting above is the quantum version of the classical broadcast channel with cooperating decoders, by Dabora and Servetto \cite{DaboraServetto:06p}. The main motivation involves a sensor network, where an external transmitter ($B_1$)
wants to download data such as network configuration into the network %
(see Ref. \onlinecite[Subsec. I A]{DaboraServetto:06p}). 
 The model can be viewed as a combination of the broadcast channel and the primitive relay channel \cite{Kim:07c}. In this context,    the term `conferencing' indicates cooperation between two different users, whereas a relay channel \cite{vanderMeulen:71p} consists of a single user and a helper (see Def. \ref{def:primRelay}).  
\end{remark}

\begin{remark}
\label{rem:bitpipe}
 The conferencing link can be described as a bit-pipe \cite{Steinberg:15c}, i.e. a noiseless link, from Receiver 1 to Receiver 2, through which information is transmitted at a constant rate $\inC_{12}$. 
\end{remark}

\begin{center}
\begin{figure}[ht!]
\includegraphics[scale=0.75,trim={1cm 10.5cm 0 11cm},clip]{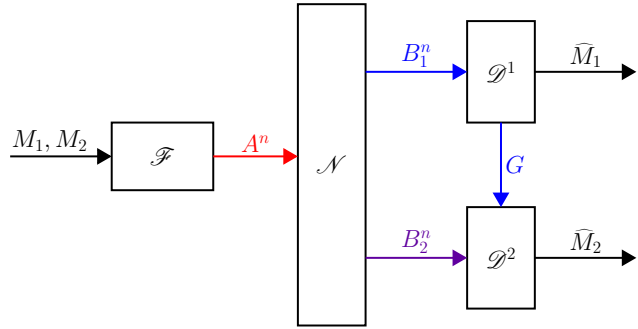} %
\caption{Quantum coding for a broadcast channel $\channel_{A\rightarrow B_1 B_2}$ with conferencing and private messages. The quantum systems of Alice, Bob 1, and Bob 2 are marked in red, blue, and purple, respectively. 
Given entanglement between the decoders, the classical conference link can be used to transfer \emph{quantum} information from Bob 1 to Bob 2 using the teleportation protocol. 
This is thus equivalent to a conferencing link with quantum capacity $\inC_{Q,12}=\frac{1}{2}\inC_{12}$.
Alice encodes the quantum state of the message systems $M_1$ and $M_2$ by applying the encoding map $\Fset_{  M_1 M_2 \rightarrow A^n}$.
Then, she transmits the systems $A^n$ over the broadcast channel. %
Bob 1 receives the channel output systems $B_1^n$ and applies the decoding map $\Dset^1_{B_1^n\rightarrow  \hM_1 G}$ such that the state of $\hM_1$ is his estimate of his private message. Next, Bob 1 sends the state of the conference system $G$ to Bob 2 through a noiseless conference link $\text{id}_{G\rightarrow G'}$. Bob 2 receives the channel output systems $B_2^n$ and the conference message in $G'$, and 
 applies the decoding map $\Dset^2_{G' B_2^n\rightarrow \hM_2 }$  such that the state of $\hM_2$ is his estimate of his private message.
}
\label{fig:BconfQ}
\end{figure}

\end{center}

\subsection{Quantum Coding with Conferencing, Entanglement Generation, and Entanglement Transmission}
\label{subsec:BcodingQF}
Next, we consider the case where the messages are quantum. Furthermore, given entanglement between the decoders, the classical conference link can be used to transfer \emph{quantum} information from Receiver 1 to Receiver 2 using the teleportation protocol. 
This is thus equivalent to a conferencing link with quantum capacity $\inC_{Q,12}=\frac{1}{2}\inC_{12}$.
In other words, given entanglement resources,  the conferencing bit-pipe of capacity $\inC_{12}$ can be transformed into a conferencing \emph{qubit}-pipe of capacity $\inC_{Q,12}=\frac{1}{2}\inC_{12}$ (see Rem.~\ref{rem:bitpipe}).
Note that due to the no-cloning theorem, the encoder cannot transmit a quantum message to both receivers, thus we consider two private messages.

\begin{definition} %
\label{def:Qcapacitycf}
A $(2^{nQ_1},2^{nQ_2},n)$ quantum code for the quantum broadcast channel $\channel_{A\rightarrow B_1 B_2}$ with
independent messages and conferencing  consists of the following: 
\begin{itemize}
\item 
A quantum message state $\rho_{M_1 M_2}$, where $M_1$ and $M_2$ are quantum systems that store the private messages of User 1 and User 2, respectively.	The dimension of each system is given by $|\Hset_{M_k}|=2^{nQ_k}$ for $k=1,2$.
\item
	an encoding map $\Fset_{M_1 M_2  \rightarrow A^n}$;
\item	
	a decoding map   $\Dset^1_{B_1^n \rightarrow \hM_1 G}  $ for Receiver 1, where $G$ is a quantum register of dimension
$2^{n\inC_{Q,12}}$	that stores the conference message from Receiver 1 to Receiver 2; 
\item
 a decoding map $\Dset^2_{G' B_2^n \rightarrow \hM_2 }  $ for Receiver 2.
\end{itemize}
We denote the code by $(\Fset,\Dset^1,\Dset^2)$.

The communication scheme is depicted in Figure~\ref{fig:BconfQ}.  
The sender Alice has the systems  $M_1$, $M_2$, and $A^n$; Bob 1 has the systems $B_1^n$,  $G$, and $\hM_1$; and
Bob 2 has the systems $B_2^n$,  $G'$, and $\hM_2$. Alice encodes the quantum state of the message systems $M_1$ and $M_2$ by applying the encoding map $\Fset_{  M_1 M_2 \rightarrow A^n}$, which results in the input state
\begin{align}
\rho_{A^n}= \Fset_{ M_1 M_2 \rightarrow A^n}( \rho_{M_1 M_2}  ) %
\end{align}
 and transmits the systems $A^n$ over %
$n$ channel uses of $\channel_{A\rightarrow B_1 B_2}$. Hence, the output state is
\begin{align}
\rho_{ B_1^n B_2^n}=\channel_{ A\rightarrow B_1 B_2}^{\otimes n} (\rho_{A^n}) \,.
\end{align}
Bob 1 receives the channel output systems $B_1^n$ and applies the decoding map $\Dset^1_{B_1^n\rightarrow  \hM_1 G}$, which results in
\begin{align}
\rho_{ \hM_1 G B_2^n}=\Dset^1_{B_1^n\rightarrow  \hM_1 G}(\rho_{ B_1^n B_2^n}) \,.
\end{align}
The reduced state of $\hM_1$ is Bob 1's estimate of the original state of his private message system $M_1$. Next, Bob 1 sends the conference message from $G$ to $G'$ using the noiseless conference link $\text{id}_{G\rightarrow G'}$,
hence $\rho_{\hM_1 G' B_2^n}=\text{id}_{G\rightarrow G'}(\rho_{\hM_1 G B_2^n})=\rho_{\hM_1 G B_2^n}$.
Bob 2 receives the channel output systems $B_2^n$ and the conference message in $G'$, and 
 applies the decoding map $\Dset^2_{G' B_2^n\rightarrow \hM_2 }$  such that $\hM_2$ is his estimate of his private message.
The estimated state is then given by
\begin{align}
\rho_{\hM_1 \hM_2 }=\Dset^2_{G' B_2^n\rightarrow \hM_2}(\rho_{\hM_1 G B_2^n}) 
\end{align}
and the estimation error by 
\begin{align}
e^{(n)}(\Eset,\Dset^1,\Dset^2,\rho_{M_1,M_2})&= 
\frac{1}{2}\norm{\rho_{M_1 M_2}-\rho_{\hM_1 \hM_2}}
\end{align}

A $(2^{nQ_1},2^{nQ_2},n,\eps)$ quantum code satisfies 
$%
 e^{(n)}(\Fset,\Dset^1,\Dset^2,\rho_{M_0 M_1})\leq\eps $ %
 for %
all $\rho_{M_0,M_1}$.  %
A rate pair $(Q_1,Q_2)$ is called achievable with conferencing  if for every $\eps>0$ and sufficiently large $n$, there exists a 
$(2^{nQ_1},2^{nQ_2},n,\eps)$ code. 
 The quantum capacity region $\opR_{\text{Q}}(\channel)$ is defined as the set of achievable pairs $(Q_1,Q_2)$ with conferencing. 
\end{definition}

\begin{center}
\begin{figure}[ht!]
\includegraphics[scale=0.5,trim={2cm 11.5cm 5cm 12cm},clip]{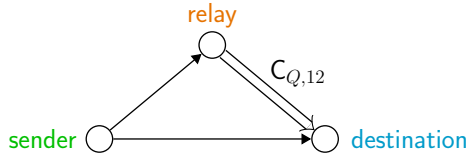} %
\caption{A simplistic view of the primitive relay channel. Alice, Bob 1, and Bob 2 play the roles of the sender, relay, and destination receiver, respectively.
}
\label{fig:QrelayP}
\end{figure}

\end{center}

The setting of a broadcast channel with conferencing is closely related to that of a primitive relay channel \cite{Kim:07c}. 
\begin{definition}
\label{def:primRelay}
A primitive relay channel $\channel^{\,\text{relay}}_{A\rightarrow B_1 B_2}$ is a broadcast channel with conferencing, when User 1 does not send information, i.e. $Q_1=0$. Alice, Bob 1, and Bob 2 are then called the source, relay, and destination receiver, respectively. A quantum rate $Q_2>0$ is called achievable for the primitive relay channel if $(0,Q_2)$ is achievable for the broadcast channel with conferencing.
The quantum capacity $C_{\text{Q}}(\channel^{\,\text{relay}})$ is defined as the supremum of achievable rates for the primitive relay channel.
\end{definition}
Bob 1 is called a relay in this setting, because his only task is to help the transmission of information to Bob 2 (see Figure~\ref{fig:QrelayP}). The channel is called `primitive' since it is a simplified version of the (non-primitive) relay channel \cite{vanderMeulen:71p} where information is received and encoded at the relay in a strictly-causal manner.

\begin{remark}
A standard, i.e. non-primitive, relay channel \cite{SavovWildeVu:12c,Savov:12z,DingGharibyanHaydenWalter:20p} is specified by a linear, completely positive, trace-preserving map 
$\mathcal{L}_{A\rightarrow B_1 A_1 B_2}$, where the sender transmits the systems $A^n$, the relay receives $B_1^n$ and transmits $A_1^n$, and the destination receiver receives $B_2^n$. The relay encoder applies a strictly-causal map, as he can only use the systems $B_1^{i-1}$ at time $i$. That is, at time $i$, the relay transmits $A_{1,i}$ such that $\rho_{A_1^i}=\mathcal{T}^{(i)}_{B_1^{i-1}\rightarrow A_1^i}(\rho_{B_1^{i-1}})$.
\end{remark}

\begin{remark}
\label{rem:EntTr}
Quantum communication is also referred to as entanglement transmission and can be extended to strong subspace transmission \cite{BjelakovicBocheNotzel:09p,AhlswedeBjelakovicBocheNotzel}.
In this task, Alice and Charlie share a pure entangled state $|\psi_{M_1 M_2 C}\rangle$, with a Schmidt decoposition
\begin{align}
|\psi_{M_1 M_2 C}\rangle=\sum_{x\in\Xset} \sqrt{p_X(x)} |\psi_{M_1 M_2}^x \rangle \otimes |\psi^x_C\rangle.
\end{align}
Given a $(2^{nQ_1},2^{nQ_2},n,\eps)$ code, Alice can send each state $ |\psi_{M_1 M_2}^x\rangle$ reliably, i.e. with  
\begin{align}
\norm{\psi_{M_1 M_2}^x-\psi_{\hM_1 \hM_2}^x}_1\leq \eps \;\text{ for all $x\in\Xset$}\,.
\end{align}
 Hence, the superposition state $|\psi_{M_1 M_2 C}\rangle$ can also be recovered up to an error of $\eps$.
In particular, if Alice and Charlie share a maximally entangled state $|\Phi_{M_1 M_2 C}\rangle$, then at the end of the communication protocol, Bob 1, Bob 2, and Charlie share a state $\approx |\Phi_{\hM_1 \hM_2 C}\rangle$ up to an $\eps$-error.
\end{remark}

\begin{remark}
Quantum communication can also be used for the purpose of entanglement generation \cite{Devetak:05p,BjelakovicBocheNotzel:09p}.
We note that by the monogamy property of quantum entanglement \cite{KoashiWinter:04p}, Alice cannot generate a maximally entangled state with both Bob 1 and Bob 2 simultaneously. %
Indeed, suppose that Alice has a third system $\bar{A}$ that is entangled with $M_1$ and $M_2$ in a state $|\psi_{\bar{A}M_1M_2}\rangle$.
Then, by strong sub-additivity (Ref. \onlinecite[Coro. 11.9.1]{Wilde:17b}),
\begin{align}
H(\bar{A} M_1)_\psi+H(\bar{A} M_1)_\psi \geq H(\bar{A})_\psi+ H(\bar{A}M_1 M_2)_\psi \geq H(\bar{A})_\psi .
\end{align}
Hence, $\bar{A}$ cannot be maximally entangled with both $M_1$ and $M_2$, otherwise we would have $0\geq 1$. 
Nevertheless, different forms of entanglement can be generated. In particular, Alice can generate a GHZ state with Bob 1 and Bob 2
(Ref. \onlinecite[Sec. IV]{YardHaydenDevetak:11p}),
 using  $|\psi_{\bar{A} M_1 M_2}\rangle=\frac{1}{\sqrt{d}}\sum_{x=1}^d |x\rangle\otimes |x\rangle\otimes |x\rangle$.
Alternatively, she can generate two entangled pairs.
Suppose that Alice has another pair of system $\bar{A}_1,\bar{A}_2$ in the state %
\begin{align}
|\psi_{\bar{A}_1 M_1 \bar{A}_2 M_2}\rangle= |\Phi_{\bar{A}_1 M_1}\rangle \otimes |\Phi_{\bar{A}_2 M_2}\rangle .
\end{align}
Then, at the end of the quantum communication protocol, Alice shares the entangled states $\approx |\Phi_{\bar{A}_1 \hM_1}\rangle$ with Bob 1 and $\approx |\Phi_{\bar{A}_2 \hM_2}\rangle$ with Bob 2, by the same considerations as in entanglement transmission (see Remark~\ref{rem:EntTr}). 
\end{remark}

\begin{remark}
In the absence of entanglement resources between the decoders, quantum communication over the broadcast channel can generate such entanglement by choosing the quantum message state to be $|\Phi_{M_1 M_2}\rangle$.
\end{remark}

\section{Main Results - Classical Conferencing}
\label{sec:Clf}
Now, we give our results on the quantum broadcast channel with a classical conferencing link between the decoders, when Bob 1 and 
Bob 2 do \emph{not} share entanglement resources (see Fig.~\ref{fig:BconfCl}).
Define the rate region
\begin{align}
\mathcal{R}_{\text{Cl}}(\channel)\triangleq \bigcup %
\left\{
\begin{array}{lrl}
(R_0,R_1) \,:\; & R_0 \leq& I(X_0;B_2)_\rho +\inC_{12} \\
								& R_1 \leq& I(X_1;B_1|X_0) \\
								& R_0+R_1 \leq& I(X_0,X_1;B_1)
\end{array}
\right\}
\label{eq:inCcl}
\end{align}
where the union is over the set of all distributions $p_{X_0,X_1}(x_0,x_1)$ and state collection $\{ \theta^{x_0,x_1}_{A}  \}$, with
\begin{align}
\label{eq:StateMaxCl}
&\rho_{X_0 X_1 B}=  \sum_{x_0\in\Xset_0}  \sum_{x_1\in\Xset_1}  p_{X_0,X_1}(x_0,x_1)  \kb{x_0} \otimes \kb{x_1} \otimes \channel_{A\rightarrow B_1 B_2}( \theta^{x_0,x_1}_{A} )
 \,. %
\end{align}
Before we state the capacity theorem, we give the following lemma which provides cardinality bounds for the auxiliary random variables
$X_0$ and $X_1$. In principle, one can use those cardinality bounds to evaluate the region $\mathcal{R}_{\text{Cl}}(\channel)$ numerically.
\begin{lemma}
\label{lemm:CardCl}
The union in (\ref{eq:inCcl}) is exhausted by auxiliary random variables $X_0$ and $X_1$ of cardinality
$
|\Xset_0|\leq |\Hset_A|^2+2
$
and
$
|\Xset_1|\leq (|\Hset_A|^2+2)|\Hset_A|^2+1				%
$.
\end{lemma}
The proof of Lemma~\ref{lemm:CardCl} is given in Appendix~\ref{app:CardCl}. The classical capacity region is determined in the theorem below.
\begin{theorem}
\label{theo:BCcl}
The classical capacity region of the quantum broadcast channel $\channel_{A\rightarrow B_1 B_2}$ with conferencing and degraded message sets is given by 
\begin{align}
\opR_{\text{Cl}}(\channel)=\lim_{k\rightarrow\infty} \bigcup_{k=1}^\infty \mathcal{R}_{\text{Cl}}(\channel^{\otimes k}) \,.
\end{align}
Furthermore, for a Hadamard broadcast channel,
\begin{align}
\opR_{\text{Cl}}(\channel)=  \mathcal{R}_{\text{Cl}}(\channel) \,.
\end{align}
\end{theorem}
The proof of Theorem~\ref{theo:BCcl} is given in Appendix~\ref{app:BCcl}.

Note that in the special case of a conference link with zero capacity, i.e. $\inC_{12}=0$, we recover the result by Yard \etal \cite{YardHaydenDevetak:08p} on the broadcast channel without conferencing.

\section{Main Results - Quantum Conferencing}
\label{sec:Qf}
Next, we consider the case where the messages are quantum. Furthermore, given entanglement between the decoders, the classical conference link can be used to transfer \emph{quantum} information from Receiver 1 to Receiver 2 using the teleportation protocol. 
This is thus equivalent to a conferencing link with quantum capacity $\inC_{Q,12}=\frac{1}{2}\inC_{12}$. 
As noted in Subsection~\ref{subsec:BcodingQF},  Alice cannot transmit a quantum message to both receivers due to the no-cloning theorem. Thereby, we consider a broadcast channel with two private quantum messages, as illustrated in Fig.~\ref{fig:BconfQ}.

This setting is intimately related to quantum repeaters, as Bob 1 can be viewed as a repeater for the transmission of quantum information to Bob 2. In particular, Alice can use the quantum message stored in $M_1$ to generate entanglement and prepare a maximally entangled pair $|\Phi_{A B_1}\rangle$ between the transmitter and the repeater, namely, Alice and Bob 1.
Given entanglement between the decoders, we also have a maximally entangled pair $|\Phi_{B_1' B_2'}\rangle$ between the repeater and the receiver, i.e. Bob 1 and Bob 2. Then, the repeater $B_1$ can swap his entanglement by using the classical conferencing link to
 teleport the state of $B_1'$ onto $B_2$ thus swapping the entanglement such that $A$ and $B_2$ are now entangled. This requires that the conferencing capacity is at least twice the information rate, i.e. $\inC_{12}\geq 2Q_2$. 
We will conclude this section with the resulting observations for the quantum repeater.

\subsection{Achievable Region}
\label{subsec:Qfachieve}
We establish an achievable rate region for the broadcast channel with quantum conferencing.
\begin{theorem}
\label{theo:BCqin}
A rate pair $(Q_1,Q_2)$ is achievable for transmission of quantum information over the broadcast channel $\channel_{A\rightarrow B_1 B_2}$ with private messages and quantum conferencing if  
\begin{align}
 Q_1 &\leq I(A_1\rangle B_1)_\rho \nonumber\\
 Q_2 &\leq I(A_2\rangle B_2)_\rho +\inC_{Q,12} \nonumber\\
 Q_1+Q_2 &\leq I(A_1\rangle B_1)_\rho +I(A_2\rangle B_2)_\rho 
\label{eq:BCqRin}
\end{align}
for some input state $\rho_{A_1 A_2 A'}$, where $\rho_{A_1 A_2 B_1 B_2}=\channel_{A'\rightarrow B_1 B_2}(\rho_{A_1 A_2 A'}) $.
\end{theorem}
The achievability proof is given below. The rate region in Theorem~\ref{theo:BCqin} reflects a greedy approach, where  using the conferencing link to increase the information rate of User 2 comes directly at the expense of User 1. That is, if $Q_2=I(A_2\rangle B_2)_\rho +\Delta$, then $Q_1\leq I(A_1\rangle B_1)_\rho -\Delta$.
\begin{remark}
For the transmission of classical information, we have seen that  the optimal performance is achieved using superposition coding, where Receiver 1 can recover the message of User 2 without necessarily ``losing" rate. In particular, by Theorem~\ref{theo:BCcl}, a classical rate pair $(R_1,R_2)=$ $(I(X_1;B_1|X_0)_\rho,I(X_0;B_2)_\rho+\inC_{12})$ is achievable when $I(X_0;B_2)_\rho +\inC_{12}<I(X_0;B_1)_\rho$, because then $R_1+R_2< I(X_0 X_1;B_1)$ by the chain rule.
 However, in the quantum case, the capacity-achieving coding scheme in Ref. \onlinecite{DupuisHaydenLi:10p} does not involve superposition.
Without conferencing, it is impossible for Receiver 1 to decode the message of User 2 by the no-cloning theorem.
Nevertheless, the setting of conferencing decoders imposes a chronological order: First Bob 1 receives and processes the channel output $B_1^n$, %
then Bob 1 sends the conference message to Bob 2, and at last, Bob 2 receives access to the channel output $B_2^n$ and the conference message. Therefore, Bob 1 \emph{can} recover the state of $M_2$  (or part of it) and send it to Bob 2 using the conference link. However, due to the no-cloning theorem, Bob 2 will be able to decode the state of $M_2$ only if the state was destroyed in Bob 1's location during conferencing. 
\end{remark}

\begin{proof}[Achievability Proof]
Consider the quantum broadcast channel $\channel_{A\rightarrow B_1 B_2}$ with a quantum conference link of capacity $\inC_{Q,12}$.
The proof is a straightforward consequence of the results by Dupuis \etal \cite{DupuisHaydenLi:10p}. Fix an input state $\rho_{A_1 A_2 A'}$.
Based on \cite{DupuisHaydenLi:10p} (Ref. \onlinecite[Theo. 5.4]{Dupuis:10z}), for every $\eps>0$ and sufficiently large $n$ there exists a
$(2^{nQ_1'},2^{nQ_2'},n,\eps)$ quantum code for the broadcast channel $\channel_{A\rightarrow B_1 B_2}$ \emph{without} conferencing if
\begin{align}
Q_1' &= I(A_1\rangle B_1)_\rho-\delta \nonumber\\
Q_2' &= I(A_2\rangle B_2)_\rho-\delta 
\label{eq:BCqRinCor1}
\end{align}
where $\delta>0$ is arbitrarily small. The rate pair $(Q_1',Q_2')$ is thus achievable in our setting as well, since the decoders can avoid conferencing by choosing an idle conference message state $\kb{0}$ regardless of the output state.

Now, we consider two cases: $\inC_{Q,12}>I(A_1\rangle B_1)_\rho$ and $\inC_{Q,12}\leq I(A_1\rangle B_1)_\rho$. If $\inC_{Q,12}>I(A_1\rangle B_1)_\rho$, then the second inequality in (\ref{eq:BCqRin}) is inactive since $I(A_1\rangle B_1)_\rho+I(A_2\rangle B_2)_\rho<
I(A_2\rangle B_2)_\rho+\inC_{Q,12}$, hence we are done. Otherwise, if $\inC_{Q,12}\leq I(A_1\rangle B_1)_\rho$, then Alice can send the state of $n\inC_{Q,12}$ qubits to Bob 2 indirectly through the conference link. This can be performed as follows.
First, use the code above to transmit $n(Q_1''+\inC_{Q,12})$ qubits to Bob 1 and $nQ_2'$ to Bob 2, with
$Q_1''=Q_1'-\inC_{Q,12}$, and then let Bob 1 send the state of the $n\inC_{Q,12}$ qubits to Bob 2. Overall, this coding scheme achieves the following rate pair, 
\begin{align}
Q_1'' &= I(A_1\rangle B_1)_\rho-\inC_{Q,12}-\delta \nonumber\\
Q_2'' &= I(A_2\rangle B_2)_\rho+\inC_{Q,12}-\delta 
\label{eq:BCqRinCor2}
\end{align}
Note that in the process of sending the conference message, Bob 1 may destroy the state of his own $n\inC_{Q,12}$ qubits, and thus this  cannot be regarded as a common message. Observing that $(Q_1',Q_2')$ and $(Q_1'',Q_2'')$ are the corner points of the region
in (\ref{eq:BCqRin}), the proof follows by time sharing.
\end{proof}
Note that for $\inC_{12}=0$, the achievable region coincides with the capacity region of the broadcast channel without conferencing \cite{DupuisHaydenLi:10p,Dupuis:10z}.

\subsection{Outer Bound}
\label{subsec:outB}
Next, we give a multi-letter outer bound. 
\begin{theorem}
\label{theo:BCqout}
If a rate pair $(Q_1,Q_2)$ is achievable for transmission of quantum information over the broadcast channel $\channel_{A\rightarrow B_1 B_2}$ with private messages and quantum conferencing, then it must satisfy the following inequalities,  
\begin{align}
 Q_1 &\leq \frac{1}{n} I(A_1\rangle B_1^n)_\rho \nonumber\\%
 Q_2 &\leq \frac{1}{n} I(A_2 T\rangle B_2^n)_\rho +\inC_{Q,12} \nonumber\\%
 Q_1+Q_2 &\leq \frac{1}{n} I(A_1 \rangle B_1^n)_\rho+\frac{1}{n} I(A_2 \rangle B_1^n B_2^n)_\rho  
\label{eq:BCqRout}
\end{align}
for some input state $\rho_{T A_1 A_2 A'^n}$, where $\rho_{T A_1 A_2 B_1^n B_2^n}=\channel^{\otimes n}_{A'\rightarrow B_1 B_2}(\rho_{T A_1 A_2 A'^n}) $.
\end{theorem}
Notice that here %
we added the auxiliary system $T$ in the second inequality and added $B_1^n$ in the last term of the third inequality (\cf (\ref{eq:BCqRin}) and (\ref{eq:BCqRout})).
\begin{proof}[Proof of Outer Bound]
Suppose that Alice is trying to generate entanglement with Bob 1 and Bob 2. %
An upper bound on the rate at which Alice and Bob $k$, for $k=1,2$, can generate entanglement also serves as an upper bound on the rate at which they can communicate qubits, since a noiseless quantum channel can be used to generate entanglement by sending one part of an entangled pair. In this task, Alice locally prepares  two maximally entangled pairs,
\begin{align}
|\Phi_{M_1 M'_1}\rangle\otimes |\Phi_{M_2 M'_2}\rangle = \frac{1}{\sqrt{2^{n(Q_1+Q_2)}}}\sum_{m_1=1}^{2^{nQ_1}} \sum_{m_2=1}^{2^{nQ_2}} | m_1 \rangle_{M_1} \otimes | m_1 \rangle_{M'_1} \otimes | m_2 \rangle_{M_2} \otimes | m_2 \rangle_{M'_2} \,.
\end{align}
 Then, she applies an encoding channel $\Fset_{M'_1 M'_2 \rightarrow A'^n}$ to the quantum systems $M'_1 M_2'$, resulting in
\begin{align}
\rho_{M_1 M_2  A'^n}\equiv \Fset_{M'_1 M'_2 \rightarrow A'^n }( |\Phi_{M_1 M'_1}\rangle\otimes |\Phi_{M_2 M'_2}\rangle ) \,.
\label{eq:QconvI1}
\end{align}
After Alice sends the systems $A'^n$ through the channel, Bob $1$ receives the systems $B_1^n$ in the state
\begin{align}
\rho_{M_1 M_2 B_1^n B_2^n}\equiv \channel^{\otimes n}_{ A'\rightarrow B_1 B_2} (\rho_{ M_1 M_2  A'^n}) %
\end{align}
and %
performs a decoding channel $\Dset^1_{B_1^n \rightarrow \hM_1 G}$. Hence, 
\begin{align}
\rho_{  M_1 M_2 \hM_1 G B_2^n}\equiv \Dset^1_{B_1^n \rightarrow \hM_1 G}(\rho_{M_1 M_2 B_1^n B_2^n})
\label{eq:DecConv1Q}
\end{align}
where the state of $\hM_1$ is Bob $1$'s estimate of his quantum message, and the state of $G$ is the conference message which is sent through the conference link to Bob $2$. Having received $B_2^n$ and $G'$ such that  $\rho_{  M_1 M_2 \hM_1 G'}=\rho_{  M_1 M_2 \hM_1 G}$,
Bob $2$ uses a decoding channel $\Dset^2_{G' B_2^n\rightarrow \hM_2}$, producing
\begin{align}
\rho_{  M_1 M_2 \hM_1 \hM_2}\equiv \Dset^2_{G' B_2^n \rightarrow \hM_2}(\rho_{M_1 M_2 \hM_1 G B_2^n}) \,.
\label{eq:DecConv2Q}
\end{align}

Consider a sequence of codes $(\Fset_n,\Dset^1_n,\Dset^2_n)$ for entanglement generation, such that
\begin{align}
\frac{1}{2} \norm{ \rho_{M_1 \hM_1 M_2 \hM_2} -\Phi_{M_1 M'_1}\otimes \Phi_{M_2 M'_2} }_1 \leq& \alpha_n \label{eq:randDconvQ} 
\end{align}
where $\alpha_n$ tends to zero as $n\rightarrow\infty$. 
By the Alicki-Fannes-Winter inequality \cite{AlickiFannes:04p,Winter:16p} (Ref. \onlinecite[Theo. 11.10.3]{Wilde:17b}), (\ref{eq:randDconvQ}) implies that $|H(M_k|\hM_k)_\rho - H(M_k|M'_k)_{\Phi} |\leq n\eps_n$, or equivalently,
\begin{align}
|I(M_k \rangle \hM_k)_\rho - I(M_k \rangle M'_k)_{\Phi} |\leq n\eps_n 
\label{eq:AFWq}
\end{align}
for $k=1,2$, where $\eps_n$ tends to zero as $n\rightarrow\infty$. 
Observe that $I(M_k \rangle M'_k)_\Phi=H(M_k)_\Phi-H(M_k M'_k)_\Phi=nQ_k-0=nQ_k$. Thus,
\begin{align}
nQ_1=& I(M_1\rangle M'_1)_\Phi \nonumber\\
  \leq& I(M_1\rangle \hM_1)_\rho+n\eps_n \nonumber\\
	\leq& I(M_1\rangle B_1^n)_\rho+n\eps_n 
	\label{eq:ineqQuna1}
\end{align}
where the last inequality is due to (\ref{eq:DecConv1Q}) and the data processing inequality for the coherent information 
(Ref. \onlinecite[Theo. 11.9.3]{Wilde:17b}).

Similarly, for User 2, it follows from (\ref{eq:DecConv2Q}) and the data processing inequality that
\begin{align}
	nQ_2%
	\leq& I(M_2\rangle G' B_2^n)_\rho+n\eps_n \label{eq:ineqQuna2s}\\
	=& -H(M_2 G'| B_2^n)+H(G'|B_2^n)+n\eps_n\nonumber\\
	=& I(M_2 G'\rangle B_2^n)+H(G'|B_2^n)+n\eps_n\nonumber\\
	\leq& I(M_2 G'\rangle B_2^n)+n\inC_{Q,12}+n\eps_n
	\label{eq:ineqQuna2}
\end{align}
By (\ref{eq:ineqQuna1}) and (\ref{eq:ineqQuna2s}), we also have that 
\begin{align}
n(Q_1+Q_2)&\leq I(M_1\rangle B_1^n)_\rho+I(M_2\rangle G' B_2^n)_\rho+2n\eps_n 
\nonumber\\
&\leq I(M_1\rangle B_1^n)_\rho+I(M_2\rangle B_1^n B_2^n)_\rho+2n\eps_n 
\label{eq:ineqQuna3}
\end{align}
where the last follows from (\ref{eq:DecConv1Q}) and the data processing inequality.
The proof follows from (\ref{eq:ineqQuna1}), (\ref{eq:ineqQuna2}) and (\ref{eq:ineqQuna3}) by defining %
 quantum systems $A_1,A_2$ such that for some isometries %
$U_{M_1\rightarrow A_1}$, $V_{M_2\rightarrow A_2}$ and $W_{ G'\rightarrow T}$, we have
$%
\rho_{A_1  B_1^n}=U_{M_1\rightarrow A_1^n}\rho_{M_1 B_1^n  } U_{M_1\rightarrow A_1^n}^\dagger 
$ and %
$%
\rho_{A_2 T B_2^n }=(V_{M_2 \rightarrow A_2}\otimes W_{ G'\rightarrow T}) \rho_{ M_2  G' B_2^n  }$ $ (V_{M_2 \rightarrow A_2}\otimes W_{ G'\rightarrow T})^\dagger %
$. %
This completes the proof for the regularized outer bound.
\end{proof}

\subsection{Primitive Relay Channel}
\label{subsec:qPrl}
Consider the primitive relay channel $\channel^{\,\text{relay}}_{A\rightarrow B_1 B_2}$, where Bob 1 acts as a relay that helps the transmission from Alice to Bob 2, but is not required to decode information (\ie $Q_1=0$). 
We use our previous results to obtain lower and upper bounds on the capacity of the primitive relay channel, and conclude this section with the resulting observations for the quantum repeater.

\begin{theorem}
\label{theo:qCpRelay}
The quantum capacity of the primitive relay channel $\channel^{\,\text{relay}}_{A\rightarrow B_1 B_2}$ has the following bounds:
\begin{enumerate}[1)]
\item
Cutset upper bound
\begin{align}
C_{\text{Q}}(\channel^{\,\text{relay}}) \leq \lim_{n\rightarrow\infty} \sup_{\rho_{ A T A'^n}} \frac{1}{n}\min\left[
I( A T\rangle B_2^n)_\rho +\inC_{Q,12} \,,\; I(A\rangle B_1^n B_2^n)_\rho \right] 
\label{eq:BCqRPinU}
\end{align}
with $\rho_{A T B_1^n B_2^n}=\channel_{A'\rightarrow B_1 B_2}^{\otimes n}(\rho_{A T A'^n}) $.

\item
Decode-forward lower bound
\begin{align}
C_{\text{Q}}(\channel^{\,\text{relay}}) \geq \max_{|\phi_{A_1 A_2 A'}\rangle} \left[
I(A_2\rangle B_2)_\rho +\min\left( I(A_1\rangle B_1)_\rho \,,\; \inC_{Q,12} \right) \right]
\label{eq:BCqRPinL}
\end{align}
with $\rho_{A_1 A_2 B_1 B_2}=\channel_{A'\rightarrow B_1 B_2}(\phi_{A_1 A_2 A'}) $.

\item
Entanglement-formation lower bound
\begin{align}
C_{\text{Q}}(\channel^{\,\text{relay}}) \geq \max_{|\phi_{A_1 A_2 A'}\rangle \,,\; \Fset_{B_1\rightarrow \widehat{B}_1} \,:\; 
E_F( \rho_{ \widehat{B}_1 A B_2 E} )\leq \inC_{Q,12}
} 
I(A_2\rangle \widehat{B}_1 B_2)_\phi
\label{eq:BCqRPcfin}
\end{align}
with $|\phi_{A B_1 B_2 E}\rangle=U^\channel_{A'\rightarrow B_1 B_2 E}|\phi_{A A'}\rangle $, 
$\rho_{A \widehat{B}_1 B_2 E}=\Fset_{B_1\rightarrow \widehat{B}_1}(\phi_{A B_1 B_2 E})$, where
$E_F( \rho_{ \widehat{B}_1 A B_2 E} )$ is the entanglement of formation with respect to the bipartition $\widehat{B}_1|AB_2 E$.
\end{enumerate}
\end{theorem}
The proof of the cutset upper bound follows the same considerations as in Subsection~\ref{subsec:outB}, and it is thus omitted (see (\ref{eq:ineqQuna2}) and (\ref{eq:ineqQuna3})). The decode-forward lower bound in Theorem~\ref{theo:qCpRelay} above is obtained as an immediate consequence of Theorem~\ref{theo:BCqin}, taking $Q_1=0$. The rate in (\ref{eq:BCqRPcfin}) can be achieved by using the conferencing link to simulate the channel $\Fset_{B_1\rightarrow \widehat{B}_1}$. Based on the results of Berta \etal (Ref. \onlinecite[Theo. 12]{BertaBradaoChristandlWehner:13p}),
this can be achieved if the capacity of the conference link is higher that the entanglement of formation with respect to the bipartition $\widehat{B}_1|AB_2 E$, i.e. 
$\inC_{Q,12}\geq E_F( \rho_{ \widehat{B}_1 A B_2 E} )$. Then, Bob 2 can decode $\rho_{\hat{B}_1^n B_2^n}$, which is $\eps_n$-close in trace distance to  
 $\rho_{\hat{B_1}^n B_2^n}\equiv \widehat{\channel}_{A\rightarrow \hat{B_1} B_2}^{\otimes n}(\rho_{A'^n})$, where $\eps_n$ tends to zero as $n\rightarrow\infty$, with 
$\widehat{\channel}_{A\rightarrow \hat{B_1} B_2}\triangleq \Fset_{B_1\rightarrow \widehat{B}_1} \circ \channel_{A\rightarrow B_1 B_2}$.

\begin{remark}
\label{rem:tradeoffP}
Recall from the beginning of Sec.~\ref{sec:Qf} that we view Alice, Bob 1, and Bob 2 as the sender, repeater, and destination receiver.
In other words, the repeater is the quantum version of a relay.
As we also consider direct transmission to the destination receiver (Bob 2), our results show the tradeoff between repeaterless communication and  relaying information through the repeater. In particular, in the decode-forward lower bound (\ref{eq:BCqRPinL}) (see part 2 of Theorem~\ref{theo:qCpRelay}), the term $I(A_2\rangle B_2)_\rho$ corresponds to repeaterless communication, while %
$\min\left( I(A_1\rangle B_1)_\rho \,,\; \inC_{Q,12}  \right)$ corresponds to quantum transmission via the repeater.
\end{remark}

\begin{remark}
\label{rem:bottlenP}
Intuitively, the decode-forward lower bound has the interpretation of %
 a bottleneck flow.
Specifically, as mentioned in the previous remark, the term $\min\left( I(A_1\rangle B_1)_\rho \,,\; \inC_{Q,12}  \right)$ in the decode-forward lower bound (\ref{eq:BCqRPinL}) is associated with the information rate via the repeater.
Due to the serial connection between the sender-repeater link $A\to B_1$ with the repeater-receiver link $B_1\to B_2$, the  throughput is dictated by the smaller rate (see Fig.~\ref{fig:QrelayP}).
A similar behavior was observed by Smolin \etal \cite{Smolin:05p} for a quantum channel with environment assistance of a classical relay (see Ref. \onlinecite[Th. 8]{Smolin:05p}, and Refs. \onlinecite{HaydenKing:04a,Winter:05a} as well).
\end{remark}

\section{Entangled Decoders}
\label{sec:EdecCl}
In this section, we consider a broadcast channel where the decoders share entanglement resources between them (see Fig.~\ref{fig:EAconfCl}). 
Given the recent results by Leditzky et al. \cite{LeditzkyAlhejjiLevinSmith:20p} on the multiple access channel, it may be tempting to think that the dual property holds for the broadcast channel and that entanglement between decoders can increase achievable rates of classical communication.
We observe that this is not the case. Nevertheless, given a quantum conferencing link of capacity $\inC_{Q,12}$, Receiver 1 can send conferencing messages to Receiver 2 at a rate $2\cdot \inC_{Q,12}$  using the super-dense coding protocol. Further details are given below.

First, consider a quantum broadcast channel $\channel_{A\to B_1 B_2}$ without conferencing, given entanglement resources shared between the decoders,  as illustrated in Fig.~\ref{fig:EAconfCl}. We show that the classical capacity region is the same as without the entanglement resources.
Indeed, suppose that Alice chooses $m_0$ and $m_1$ uniformly at random, and prepares an input state $\rho^{m_0,m_1}_{ A^n}$. 
After Alice sends the systems $A^n$ through the channel,  the output state is $\rho_{B_1^n B_2^n}\otimes \Psi_{S_{B_1} S_{B_2}}$,
where $S_{B_1}$ and $S_{B_2}$ are the entangled systems of Bob 1 and Bob 2, respectively, and
$\rho_{B_1^n B_2^n}=\frac{1}{2^{n(R_0+R_1)}}\sum_{m_0,m_1} \channel_{A^n\rightarrow B_1^n B_2^n}(\rho^{m_0,m_1}_{ A^n}) $.
Then, Bob 1 performs a decoding POVM $\Lambda^{m_0,m_1}_{B_1^n S_{B_1}}$, and Bob 2 performs a decoding POVM $\Lambda^{m_0}_{B_2^n S_{B_2}}$.
Consider a sequence of codes $(\Fset_n,\Lambda_n,\Gamma_n)$ such that the average probability of error tends to zero, hence
the error probabilities $\prob{ \hM_0\neq M_0 }$, $\prob{ (\hM_0,\hM_1)\neq (M_0,M_1)}$, $\prob{ \hM_1\neq M_1 |M_0}$ are bounded by some
$\alpha_n$ which tends to zero as $n\rightarrow \infty$.
By Fano's inequality, it follows that%
\begin{align}
H(M_0|\tM_0) \leq n\eps_n\\
H(M_0,M_1|\hM_0,\hM_1) \leq n\eps_n'\\
H(M_1|\hM_1,M_0) \leq n\eps_n''
\end{align}
where $\eps_n,\eps_n',\eps_n''$ tend to zero as $n\rightarrow\infty$.
Hence, 
\begin{align}
nR_0= H(M_0)=I(M_0;\tM_0)_{\rho}+H(M_0|\tM_0) &\leq I(M_0;\tM_0)_{\rho}+n\eps_n \nonumber\\
&\leq I(M_0;B_2^n S_{B_2})_{\rho}+n\eps_n= I(M_0;B_2^n)_{\rho}
\label{eq:ConvIneq1SCe}
\intertext{where the second inequality %
follows from the Holevo bound (see Ref. \onlinecite[Theo. 12.1]{NielsenChuang:02b}), and the last inequality holds as $S_{B_1}S_{B_2}$ are in a product state with $M_0,M_1,B_1^n,B_2^n$. Similarly,}
n(R_0+R_1)&\leq I(M_0,M_1;B_1^n)_{\rho}+n\eps_n' \label{eq:ConvIneq2SCeA} \\
nR_1&\leq I(M_1;B_1^n|M_0)_{\rho}+n\eps_n' .
\label{eq:ConvIneq2SCeB}
\end{align}
as without entanglement resources.

We observe that the this property can be extended to any pair of non-signaling correlated resources that are shared between the decoders. Specifically, suppose that Bob 1 and Bob 2 have random elements $\beta_1$ and $\beta_2$, that follow a non-signaling correlation $(\beta_1,\beta_2)\sim p(b_1,b_2|x_1,x_2)$, such that
\begin{align}
\sum_{b_2} p(b_1,b_2|x_1,x_2)&= p(b_1|x_1) \\
\sum_{b_1} p(b_1,b_2|x_1,x_2)&= p(b_2|x_2)
\end{align} 
for some $x_1$ and $x_2$. We think of $\beta_k$  as the measurement outcome of Bob $k$, while $x_k$ is his choice of measurement, for $k=1,2$.
The derivation of the outer bound above boils down to the fact that Bob 1 and Bob 2's resources are uncorrelated with the message and the channel outputs. 
Thus, the capacity region of the broadcast channel with non-signaling correlated resources between the decoders is the same as without those resources.

On the other hand, given a quantum conferencing link of capacity $\inC_{Q,12}$, the classical capacity region with entanglement between the decoders is given by the regularization of the region in (\ref{eq:inCcl}), taking $\inC_{12}=2\cdot\inC_{Q,12}$.
Achievability follows by using the super-dense coding protocol \cite{BennetWiesner:92p} to send classical conferencing messages from Bob 1 to Bob 2. As for the converse proof, consider a coding scheme where Bob 1 performs a decoding POVM $\Lambda^{m_0,m_1,g}_{B_1^n S_{B_1}}$, sends $g$ to Bob 2 using conferencing, and Bob 2 chooses a POVM
$\Gamma^{m_0}_{B_2^n S_{B_2}|g}$ accordingly. By the same considerations as in the derivation above,
\begin{align}
nR_0%
&\leq I(M_0;\tM_0)_{\rho}+n\eps_n \nonumber\\
&\leq I(M_0;B_2^n S_{B_2} G')_{\rho}+n\eps_n \nonumber\\
&= I(M_0;B_2^n S_{B_2})_{\rho} + I(M_0;G'|B_2^n S_{B_2})_{\rho}+n\eps_n  \nonumber\\
&\leq I(M_0;B_2^n)_{\rho} + 2\cdot n\inC_{Q,12}+n\eps_n 
\label{eq:ConvIneq1SCeQf}
\end{align}
where the second inequality %
follows from the Holevo bound (see Ref. \onlinecite[Theo. 12.1]{NielsenChuang:02b}), the equality is due to chain rule for the quantum mutual information, and the last inequality holds because $I(M_0;B_2^n S_{B_2})_{\rho}=I(M_0;B_2^n)_{\rho}$
 as $S_{B_1}S_{B_2}$ are in a product state with $M_0,M_1,B_1^n,B_2^n$, and since $I(M_0;G'|B_2^n S_{B_2})_{\rho}\leq 2H(G')_\rho\leq 2\cdot n\inC_{Q,12}$ (see Ref. \onlinecite[Sec. 11.6]{Wilde:17b}). As the bounds (\ref{eq:ConvIneq2SCeA})-(\ref{eq:ConvIneq2SCeB}) hold by similar arguments, the  proof follows.

We conclude that entanglement between the decoders cannot enlarge the capacity region of the classical broadcast channel without conferencing. By similar considerations, the same property holds for a broadcast channel with classical conferencing as well, and more generally, for any pair of non-signaling correlated resources. Yet, entanglement resources between the decoders double the conferencing rate when a quantum conferencing link is available. 
Further observations and a comparison with the multiple access channel are provided in the discussion section, in Subsec.~\ref{subsec:duality}.

\section{Summary and Discussion}
\label{sec:summary}
We have considered the quantum broadcast channel $\channel_{A\rightarrow B_1 B_2}$ in different settings of cooperation between the decoders. Using those settings, we provided an information-theoretic framework for quantum repeaters.

\subsection{Conferencing}
\label{subsec:conferencing}
The first form of cooperation that we considered is classical conferencing, where Receiver 1 can send classical messages to Receiver 2.  
We provided a regularized characterization for the classical capacity region of the quantum broadcast channel with classical conferencing, and a single-letter formula for Hadamard broadcast channels.
Next, we considered quantum conferencing, where Receiver 1 can teleport a quantum state to Receiver 2.
We developed inner and outer bounds on the quantum capacity region with quantum conferencing, characterizing the tradeoff between the communication rates $Q_1$ and $Q_2$ to Receiver 1 and Reciever 2, respectively, as well as the conferencing capacity $\inC_{Q,12}$.

Quantum communication is also referred to as entanglement transmission and can be extended to strong subspace transmission \cite{BjelakovicBocheNotzel:09p,AhlswedeBjelakovicBocheNotzel}.
In this task, Alice and Charlie share a pure entangled state $|\psi_{M_1 M_2 C}\rangle$, and
at the end of the communication protocol, Bob 1, Bob 2, and Charlie share a state $\approx |\Phi_{\hM_1 \hM_2 C}\rangle$ up to an $\eps$-error.
In the absence of entanglement resources between the decoders, quantum communication over the broadcast channel can generate such entanglement by choosing the quantum message state to be $|\Phi_{M_1 M_2}\rangle$.

Quantum communication can also be used for the purpose of entanglement generation \cite{Devetak:05p,BjelakovicBocheNotzel:09p}.
We note that by the monogamy property of quantum entanglement \cite{KoashiWinter:04p}, Alice cannot generate a maximally entangled state with both Bob 1 and Bob 2 simultaneously. %
Nevertheless, different forms of entanglement can be generated. In particular, Alice can generate a GHZ state with Bob 1 and Bob 2
(Ref. \onlinecite[Sec. IV]{YardHaydenDevetak:11p}),
 using  $|\psi_{\bar{A} M_1 M_2}\rangle=\frac{1}{\sqrt{d}}\sum_{x=1}^d |x\rangle\otimes |x\rangle\otimes |x\rangle$.
Alternatively, she can generate two entangled pairs.
Suppose that Alice has another pair of system $\bar{A}_1,\bar{A}_2$ in the state %
\begin{align}
|\psi_{\bar{A}_1 M_1 \bar{A}_1 M_2}\rangle= |\Phi_{\bar{A}_1 M_1}\rangle \otimes |\Phi_{\bar{A}_2 M_2}\rangle .
\end{align}
Then, at the end of the quantum communication protocol, Alice shares the entangled states $\approx |\Phi_{\bar{A}_1 \hM_1}\rangle$ with Bob 1 and $\approx |\Phi_{\bar{A}_2 \hM_2}\rangle$ with Bob 2. %

The case where Receiver 1 is not required to recover information, i.e. $Q_1=0$, and its sole purpose is to help the transmission to Receiver 2, reduces to the quantum primitive relay channel, for which the decode-forward lower bound and cutset upper bound follow as a consequence. In addition, we established an entanglement-formation lower bound,  
where a virtual channel is simulated through the conference link, following the results of Berta \etal \cite{BertaBradaoChristandlWehner:13p} on quantum channel simulation.

\subsection{Quantum Repeaters}
\label{subsec:repeaters}
  The quantum conferencing setting is intimately related to quantum repeaters, as the sender, Receiver 1, and Receiver 2 can be viewed as the transmitter, the repeater, and the destination receiver, respectively, in the  repeater model. In particular, the sender can employ quantum communication to Receiver 1 (the repeater)  in order to prepare a maximally entangled pair $|\Phi_{A B_1}\rangle$, which consists of $nQ_1$ entangled bits (ebits).

Given entanglement between the receivers, we also have a maximally entangled pair $|\Phi_{B_1' B_2'}\rangle$, which consists of $n\inC_{Q,12}=\frac{1}{2}n\inC_{12}$ ebits, shared between the repeater and the destination receiver, where $\inC_{12}$ is the classical conferencing link.  
 Then, the repeater can swap his entanglement by using the classical conferencing link to
 teleport the state of $B_1'$ onto $B_2$ thus swapping the entanglement such that $A$ and $B_2$ are now entangled.
This requires the classical conferencing rate to be at least twice the information transmission rate to $B_2$.
 
Hence our results provide an information-theoretic analysis characterizing the achievable rates of ebits that can be generated in each stage.
As we have also considered direct transmission to the destination receiver, our results reflect the tradeoff between repeaterless communication and  relaying qubits using the repeater as well (see Rem.~\ref{rem:tradeoffP}). 

Intuitively, the communication via the repeater gives rise to a bottleneck effect.
That is,  due to the serial connection between the sender-repeater link $A\to B_1$ with the repeater-receiver link $B_1\to B_2$, the  throughput is dictated by the smaller rate (see Fig.~\ref{fig:QrelayP}).
 Indeed, the term in the decode-forward formula (\ref{eq:BCqRPinL}) %
that is associated with communication via the repeater involves a minimum between the coherent information $I(A_1\rangle B_1)_\rho$ and the conferencing link capacity $\inC_{Q,12}$ (see Rem.~\ref{rem:bottlenP}).

\subsection{BC-MAC Duality}
\label{subsec:duality}
The duality between the broadcast channel and the multiple access channel (BC-MAC duality) is a well-known property in the study of Gaussian multiple-input multiple-output (MIMO) channels \cite{JindalVishwanathGoldsmith:04p,VishwanathTse:03p,WeingartenSteinbergShamai:06p1}. 
Based on the reciprocity property \cite{Telatar:99p}, the capacity remains unchanged when the role of the transmitters and receivers is interchanged \cite{JindalVishwanathGoldsmith:04p,VishwanathTse:03p,WeingartenSteinbergShamai:06p1} (see also Ref. \onlinecite[Lemm. 9.2]{ElGamalKim:11b}). In the scalar case, %
this means that the capacity region of the Gaussian broadcast channel,
\begin{align}
&Y_1=h_1X+Z_1 \\
&Y_2=h_2 X+Z_2
\end{align}
subject to a power constraint $\frac{1}{n}\sum_{i=1}^n x_i^2\leq P$, is exactly the same as the capacity region of the Gaussian
multiple access channel,
\begin{align}
Y=h_1 X_1 +h_2 X_2 +Z
\end{align}
subject to a total-power constraint $\frac{1}{n}\sum_{i=1}^n (x_{1,i}^2+x_{2,i}^2)\leq P$, with normalized Gaussian noise $Z$, $Z_1$, $Z_2$ $\sim \mathcal{N}(0,1)$. 
As the multiple access channel and broadcast channel are useful models for uplink and downlink transmission in cellular communication, 
this behavior is also referred to as \emph{uplink-downlink duality}.
 Duality properties have also been shown for beamforming strategies \cite{FarrokhiLiuTassiulas:98p,BocheSchubert:02c}.

\begin{center}
\begin{figure}[ht!]
\includegraphics[scale=0.75,trim={2cm 9cm -1cm 9cm},clip]{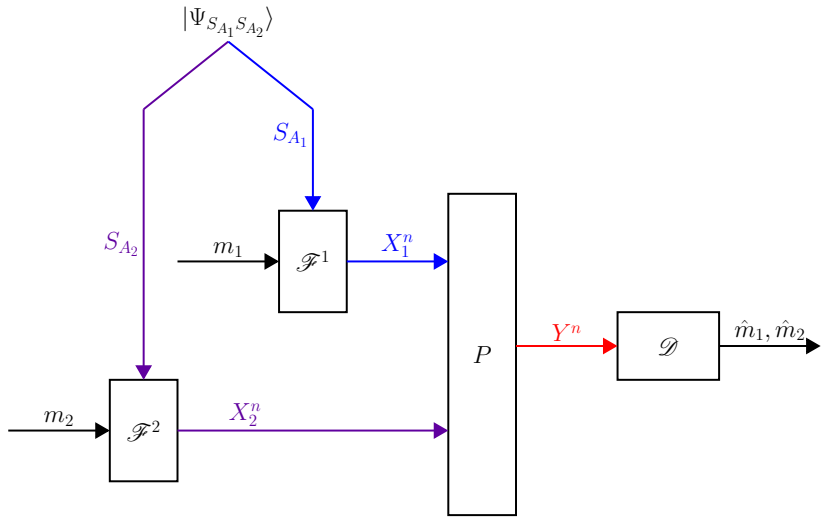} %
\caption{Classical coding for a classical multiple access channel $P_{Y|X_1 X_2}$ with shared entanglement between the encoders. The systems of Alice 1, Alice 2, and Bob are marked in  blue, purple and red, respectively. 
Alice 1 and Alice 2 share entanglement resources in the systems $S_{A_1}$ and $S_{A_2}$, respectively.
For $k=1,2$, Alice $k$ encodes the message $m_k$ by applying an encoding map $\Fset_{  M_k \rightarrow x_k^n}$ to the register $M_k$ which stores the  respective message. 
Then, she transmits $X_k^n$ over the multiple access channel. %
Bob receives the channel output sequence $Y^n$, and estimates the messages using a decoding map $\Dset_{Y^n\rightarrow \hM_1 \hM_2}$. 
Leditzky \etal \cite{LeditzkyAlhejjiLevinSmith:20p} showed that using the entanglement resources, $S_{A_1}$ and $S_{A_2}$, can strictly increase the achievable rates. We have shown that for the broadcast dual in Fig.~\ref{fig:EAconfCl}, entanglement resources cannot increase the achievable rates.
}
\label{fig:EAconfClmac}
\end{figure}

\end{center}

Our result demonstrates the limitations of the duality between the broadcast channel and the multiple access channel.
Leditzky \etal \cite{LeditzkyAlhejjiLevinSmith:20p} considered a classical multiple access channel $P_{Y|X_1 X_2}$, with two senders and a single receiver, when the encoders share entanglement resources, as illustrated in Fig.~\ref{fig:EAconfClmac}.
The MAC in Ref. \onlinecite{LeditzkyAlhejjiLevinSmith:20p} is defined in terms of a  pseudo-telepathy game \cite{BrassardBroadbentTapp:05p}, for which %
quantum strategies guarantee a certain win and outperform classical strategies. 
They showed achievability of a sum-rate $R_1+R_2$ that exceeds the sum-rate capacity of this channel without entanglement.
In principle, one could mirror the model (\cf Fig.~\ref{fig:EAconfCl} and Fig.~\ref{fig:EAconfClmac}), and consider a broadcast channel $P_{Y_1' Y_2'|X'}$, where $X'\equiv Y$, $Y_1\equiv X_1$, and 
$Y_2'\equiv X_2$, according to the a posteriori probability distribution $P_{X_1 X_2| Y}$ given some input distribution $p_{X_1,X_2}$.  %
Specifically, the derivation in Ref. \onlinecite{LeditzkyAlhejjiLevinSmith:20p} is for the magic square game \cite{BrassardBroadbentTapp:05p}, which is highly symmetric. %
Hence, it can be shown that the sum-rate capacity of the multiple access channel $P_{Y|X_1 X_2}$ and the broadcast channel $P_{Y_1' Y_2'|X'}$, without entanglement resources, are the same. Nevertheless, we cannot use the entanglement cooperation in the same manner, as the decoding strategy does not affect the channel.
Intuitively, encoding using quantum game strategies for the multiple access channel inserts quantum correlations into the channel. On the other hand, in the broadcast setting, the entangled resources of the decoders are not correlated with the channel inputs or outputs. %
This observation explains the asymmetry with regard to entanglement cooperation, and more generally, for %
any pair of non-signaling correlated resources that are shared between the decoders. Therefore, our result reveals a fundamental asymmetry and demonstrates the limitations of the duality between the broadcast channel and the multiple access channel.

\begin{acknowledgments}
Uzi Pereg, Christian Deppe, and Holger Boche were supported by the Bundesministerium f\"ur Bildung und Forschung (BMBF) through Grants
16KIS0856 (Pereg, Deppe), 16KIS0858 (Boche), and the Israel CHE Fellowship for Quantum Science and Technology (Pereg).
This work of H. Boche was also supported in part
by the German Research Foundation (DFG),
within the Gottfried Wilhelm Leibniz Prize under Grant BO 1734/20-1 and
within Germany's Excellence Strategy EXC-2111—390814868.
\end{acknowledgments}

\section*{Data Availability}
The data that supports the findings of this study are available within the article.

\appendix

\section{Proof of Lemma~\ref{lemm:CardCl}}
\label{app:CardCl}
To bound the alphabet size of the random variables $X_0$ and $X_1$, we use the Fenchel-Eggleston-Carath\'eodory lemma \cite{Eggleston:66p} and similar arguments as in Refs. \onlinecite{YardHaydenDevetak:08p,
Pereg:19a3}.
Let
\begin{align}
L_0=&|\Hset_A|^2+2 \label{eq:L0} \\
L_1=& L_0|\Hset_A|^2+1 \,. \label{eq:L1}
\end{align}

First, fix $p_{X_1|X_0}(x_1|x_0)$, and consider the ensemble $\{ p_{X_0}(x_0)p_{X_1|X_0}(x_1|x_0) \,, \theta_A^{x_0,x_1}  \}$. 
Every pure state $\theta_A=\kb{ \phi_A }$ has a unique parametric representation $u(\theta_A)$ of dimension $|\Hset_A|^2-1$. 
Then, define a map $f_0:\Xset_0\rightarrow \mathbb{R}^{L_0}$ by
\begin{align}
f_0(x_0)= \left(  u(\rho_A^{x_0}) \,,\; H(B_2|X_0=x_0)   \,,\; H(B_1|X_0=x_0)_\rho  \,,\; H(B_1|X_0=x_0,X_1)  \right)
\end{align}
where $\rho_A^{x_0}=\sum_{x_1}  p_{X_1|X_0}(x_1|x_0) \theta_A^{x_0,x_1}$. The map $f_0$ can be extended to a map that  acts on probability distributions as follows,
\begin{align}
F_0 \,:\; p_{X_0}  \mapsto
\sum_{x_0\in\Xset_0} p_{X_0}(x_0) f_0(x_0)= \left(  u(\rho_A) \,,\; H(B_2|X_0)  \,,\; H(B_1|X_0)_\rho  \,,\; H(B_1|X_0,X_1)    \right) %
\end{align}
where $\rho_A=\sum_{x_0} p_{X_0}(x_0) \rho_A^{x_0}$.
According to the Fenchel-Eggleston-Carath\'eodory lemma \cite{Eggleston:66p}, any point in the convex closure of a connected compact set within $\mathbb{R}^d$ belongs to the convex hull of $d$ points in the set. 
Since the map $F_0$ is linear, it maps  the set of distributions on $\Xset_0$ to a connected compact set in $\mathbb{R}^{L_0}$, where $L_0=(|\Hset_A|^2-1)+1+1=|\Hset_A|^2+1$  as defined in  (\ref{eq:L0}). Thus, for every  $p_{X_0}$, 
there exists a probability distribution $p_{\bar{X}_0}$ on a subset $\bar{\Xset}_0\subseteq \Xset_0$ of size $%
L_0$, such that 
$%
F_0(p_{\bar{X}_0})=F_0(p_{X_0}) %
$. %
We deduce that alphabet size can be restricted to $|\Xset_0|\leq L_0$, while preserving $\rho_A$ and
$\rho_{B_1 B_2}\equiv \channel_{A\rightarrow B_1 B_2}(\rho_A)$; $I(X_0;B_2)_\rho=H(B_2)_\rho-H(B_2|X_0)_\rho$, 
 $I(X_1;B_1|X_0)_\rho=H(B_1|X_0)_\rho-H(B_1|X_0,X_1)_\rho$; and
$I(X_0,X_1;B_1)_\rho=H(B_1)_\rho-H(B_1|X_0,X_1)_\rho$.

We move to the alphabet size of $X_1$. Fix $p_{X_0|X_1}$, where
\begin{align}
p_{X_0|X_1}(x_0|x_1)\equiv \frac{  p_{X_0}(x_0) p_{X_1|X_0}(x_1|x_0)}{   \sum_{x_0'\in\Xset_0} p_{X_0}(x_0') p_{X_1|X_0}(x_1|x_0')}
\,.
\end{align}
Define the map $f_1:\Xset_1\rightarrow \mathbb{R}^{L_1}$ by
\begin{align}
f_{1}(x_1)=& \left( p_{X_0|X_1}(\cdot|x_1) \,,\; (u(\rho_A^{x_0,x_1}))_{x_0\in\Xset_0} \,,\; H(B_1|X_0,X_1=x_1)_{\rho}   \right) %
\end{align}
where  $\rho_A^{x_1}=\sum_{x_0} p_{X_0|X_1}(x_0|x_1) \theta_A^{x_0,x_1}$. Now, the extended map is
\begin{align}
F_1 \,:\; p_{X_1} \mapsto  \sum_{x_1\in\Xset_1} p_{X_1}(x_1) f_1(x_1) = \left( p_{X_0} \,,\;  (u(\rho_A^{x_0}))_{x_0\in\Xset_0}  \,,\; 
H(B_1|X_0,X_1)_{\rho}      \right) \,.
\end{align}
By the Fenchel-Eggleston-Carath\'eodory lemma \cite{Eggleston:66p}, for every  $p_{X_1}$, 
there exists $p_{\bar{X}_1}$ on a subset $\bar{\Xset}_1\subseteq \Xset_1$ of size $(|\Hset_A|^2-1)L_0+2\leq L_1 $ (see (\ref{eq:L0})), such that 
$F_1(p_{\bar{X}_1})=F_1(p_{X_1})$.
We deduce that alphabet size can be restricted to $|\Xset_1|\leq L_1 $, while preserving $\rho_A^{x_0}$, $\rho_A$ and $\rho^{x_0}_{B_1 B_2}\equiv \channel_{A\rightarrow B_1 B_2}(\rho_A^{x_0})$,
$\rho_{B_1 B_2}\equiv \channel_{A\rightarrow B_1 B_2}(\rho_A)$; 
 $I(X_1;B_1|X_0)_\rho=H(B_1|X_0)_\rho-H(B_1|X_0,X_1)_\rho$; and
$I(X_0,X_1;B_1)_\rho=H(B_1)_\rho$ $-H(B_1|X_0,X_1)_\rho$.
\qed

\section{Proof of Theorem~\ref{theo:BCcl}}
\label{app:BCcl}
Consider a  quantum broadcast channel $\channel_{A\rightarrow B_1 B_2}$ with classical conferencing link of capacity $\inC_{12}$.
The proof extends techniques that were used in a previous work by the first author \cite{Pereg:19a3,Pereg:20c1}.

\subsection{Achievability Proof}

We show that for every $\zeta_0,\zeta_1,\eps_0>0$, there exists a $(2^{n(R_0-\zeta_0)},2^{n(R_1-\zeta_1)},n,\eps_0)$ code for 
$\channel_{A\rightarrow B_1 B_2}$ with conferencing and degraded message sets, provided that $(R_0,R_1)\in \mathcal{R}_{\text{Cl}}(\channel)$. 
To prove achievability, we extend  the classical superposition coding with binning technique to the quantum setting, and then apply the quantum packing lemma. Similar observations as in Refs. \onlinecite{Pereg:19a3,Pereg:20c1} are used as well.
Let $\{ p_{X_0}(x_0)p_{X_1|X_0}(x_1|x_0) , \theta_{A}^{x_0,x_1} \}$ be a  given ensemble, and define 
\begin{align}
\rho_{B_1,B_2}^{x_0,x_1}&\equiv \channel_{A\rightarrow B_1 B_2}(\theta_A^{x_0,x_1})\\
\sigma_{B_2}^{x_0}&\equiv \sum_{x_1'\in\Xset_1} p_{X_1|X_0}(x_1'|x_0) \rho_{B_2}^{x_0,x_1}
\end{align}
for $(x_0,x_1)\in\Xset_0\times\Xset_1$, where $\rho_{B_2}^{x_0,x_1}$ is the reduced state of $\rho_{B_1,B_2}^{x_0,x_1}$.

Standard method-of-types concepts are defined as in Refs. \onlinecite{Wilde:17b,Pereg:19a3}.  
We briefly introduce the notation and basic properties while the detailed definitions can be found in Ref. \onlinecite[Sec. III]{Pereg:19a3}.
In particular, given a density operator $\rho=\sum_x p_X(x)\kb{x}$ on the Hilbert space $\Hset_A$, we let
$\tset(p_X)$ denote the $\delta$-typical set that is associated with $p_X$, and
 $\Pi_{A^n}^{\delta}(\rho)$ the projector onto the corresponding subspace.  
The following inequalities follow from well-known properties of $\delta$-typical sets \cite{NielsenChuang:02b}, %
\begin{align}
\trace( \Pi^\delta(\rho) \rho^{\otimes n} )\geq& 1-\eps  \label{eq:UnitT} \\
 2^{-n(H(\rho)+c\delta)} \Pi^\delta(\rho) \preceq& \,\Pi^\delta(\rho) \,\rho^{\otimes n}\, \Pi^\delta(\rho) \,
\preceq 2^{-n(H(\rho)-c\delta)}
\label{eq:rhonProjIneq}
\\
\trace( \Pi^\delta(\rho))\leq& 2^{n(H(\rho)+c\delta)} \label{eq:Pidim}
\end{align}
 where $c>0$ is a constant.
Furthermore, for $\sigma_B=\sum_x p_X(x)\rho_B^x$, %
let $\Pi_{B^n}^{\delta}(\sigma_B|x^n)$ denote the projector corresponding to the conditional $\delta$-typical set %
given the sequence $x^n$.
Similarly \cite{Wilde:17b}, %
\begin{align}
\trace( \Pi^\delta(\sigma_B|x^n) \rho_{B^n}^{x^ n} )\geq& 1-\eps'  \label{eq:UnitTCond} \\
 2^{-n(H(B|X')_\sigma+c'\delta)} \Pi^\delta(\sigma_B|x^n) \preceq& \,\Pi^\delta(\sigma_B|x^n) \,\rho_{B^n}^{x^ n}\, \Pi^\delta(\sigma_B|x^n) \,
\preceq 2^{-n(H(B|X')_{\sigma}-c'\delta)}
\label{eq:rhonProjIneqCond}
\\
\trace( \Pi^\delta(\sigma_B|x^n))\leq& 2^{n(H(B|X')_\sigma+c'\delta)} \label{eq:PidimCond}
\end{align}
where $c'>0$ is a constant, $\rho_{B^n}^{x^n}=\bigotimes_{i=1}^n \rho_{B_i}^{x_i}$, and the classical random variable $X'$ is distributed according to the type of $x^n$.
If $x^n\in\tset(p_X)$, then %
\begin{align}
\trace( \Pi^\delta(\sigma_B) \rho_{B^n}^{x^n} )\geq& 1-\eps' \,. 
\label{eq:UnitTCondB}
\end{align}
 as well (see Ref. \onlinecite[Property 15.2.7]{Wilde:17b}).
We note that the conditional entropy in the bounds above can also be expressed as 
$%
H(B|X')_\sigma=\frac{1}{n} H(B^n|X^n=x^n)_{\sigma}  \equiv \frac{1}{n} H(B^n)_{\rho^{x^n}} %
$. %

The code construction, encoding and decoding procedures are described below.

\emph{Classical Code Construction:} 
Select $2^{nR_0}$ independent sequences $x_0^n(m_0)$, $m_0\in [1:2^{nR_0}]$, 
at random according to $\prod_{i=1}^n p_{X_0}(x_{0,i})$. For every $m_0\in [1:2^{nR_0}]$, select $2^{n R_1}$ conditionally independent sequences $x_1^n(m_0,m_1)$, $m_1\in [1:2^{n R_1}]$,
at random according to $\prod_{i=1}^n p_{X_1|X_0}(x_{1,i}|x_{0,i}(m_0)$. 
Partition the set of indices $[1:2^{nR_0}]$ into $2^{n\inC_{12}}$ bins of equal size,
\begin{align}
\mathscr{B}(g)=[(g-1)2^{n(R_0-\inC_{12})}: g 2^{n(R_0-\inC_{12})}]
\label{ref:bindef} 
\end{align}
for $g\in [1:2^{\inC_{12}}]$.

\emph{Encoding:}
To send the message pair $(m_0,m_1)$,  Alice prepares $\rho_{A^n}= \bigotimes_{i=1}^n \rho_A^{x_{1,i}(m_0,m_1)  }$ and sends the block $A^n$. 
The resulting output state is
\begin{align}
\rho_{B_1^n B_2^n}= \bigotimes_{i=1}^n \rho_B^{x_{0,i}(m_0),x_{1,i}(m_0,m_1)}
\end{align}

\emph{Decoding:}
Bob 1 receives the systems $B_1^n$ and decodes as follows. 
\begin{enumerate}[(i)]
\item
Decode $\htm_0$ by applying a POVM $\{ \Lambda^0_{m_0} \}_{
m_0 \in  [1:2^{nR_0}]}$,  to the systems $B_1^n$. 
\item
Decode $\htm_1$ by applying a second POVM $\{ \Lambda^1_{m_1|\htm_0} \}_{
m_1 \in  [1:2^{nR_1}]}$,  to the systems $B_1^n$.
\item
Choose $g$ to be the corresponding bin index such that $\htm_0\in\mathscr{B}(g)$. 
\item
Send the conference message $g$ to Bob 2.
\end{enumerate}
where the POVMs $\{ \Lambda^0_{m_0} \}$ and $\{ \Lambda^1_{m_1|\htm_0} \}$ will be specified later.

Bob 2 receives the systems $B_2^n$ and the conference message $g$ and decodes by applying a POVM $\{ \Gamma_{m_0} \}_{
m_0 \in  \mathscr{B}(g)}$, which will also be specified later, to the systems $B_2^n$.

\emph{Analysis of Probability of Error:}
Assume without loss of generality that Alice sends $(m_0,m_1)$. Denote the decoding measurement outcomes by
$\hM_0$, $\hM_1$, $G$, and $\tM_0$.
Consider the following events,
\begin{align}
\mathscr{E}_1=& \{  (X_0^n(m_0),X_1^n(m_0,m_1))\notin \Aset^{\delta/2}(p_{X_0,X_1})   \} \\
\mathscr{E}_2=& \{  \hM_0\neq m_0  \}\\
\mathscr{E}_3=& \{  \hM_0= m_0,\hM_1\neq m_1  \}\\
\mathscr{E}_4=& \{  \tM_0 \neq m_0  \} 
\end{align}
By the union of events bound, the probability of error is bounded by
\begin{align}
P_{e|m_0,m_1}^{(n)}(\Fset,\Lambda,\Gamma) \leq& %
 \prob{ \mathscr{E}_1 }%
+ \cprob{ \mathscr{E}_2 }{ \mathscr{E}_1^c} 
+\cprob{ \mathscr{E}_3 }{ \mathscr{E}_1^c\cap  \mathscr{E}_2^c } 
+\cprob{ \mathscr{E}_4 }{ \mathscr{E}_1^c \cap \mathscr{E}_2^c \cap \mathscr{E}_3^c } \,.
\label{eq:PeBsc}
\end{align}
The first term tends to zero as $n\rightarrow\infty$ by the law of large numbers. 
To bound the second term, we use the quantum packing lemma. Given $\mathscr{E}_1^c$, we have $(X_0^n(m_0),X_1^n(m_0,m_1))\in\Aset^{
\nicefrac{\delta}{2}}(p_{X_0,X_1})$. %
Now, observe that
\begin{align}
\Pi^{\delta}(\rho_{B_1})  \rho_{B_1^n}   \Pi^{\delta}(\rho_{B_1}) \preceq& 2^{ -n(H(B_1)_{\rho}-\eps_2(\delta)) } \Pi^{\delta}(\rho_{B_1})
\\
\trace\left[ \Pi^{\delta}(\rho_{B_1}|x_0^n,x_1^n) \rho_{{B_1}^n}^{x_0^n,x_1^n} \right] \geq& 1-\eps_2(\delta) \\
\trace\left[ \Pi^{\delta}(\rho_{B_1}|x_0^n,x_1^n)  \right] \leq& 2^{ n(H({B_1}|X_0,X_1)_{\rho} +\eps_2(\delta))} \\
\trace\left[ \Pi^{\delta}(\rho_{B_1}) \rho_{{B_1}^n}^{x_0^n,x_1^n} \right] \geq& 1-\eps_2(\delta) 
\end{align}
for all $(x_0^n,x_1^n)\in\Aset^{\delta/2}(p_{X_0,X_1})$, by (\ref{eq:rhonProjIneq}), (\ref{eq:UnitTCond}),  (\ref{eq:PidimCond}), and 
(\ref{eq:UnitTCondB}), respectively. By the quantum packing lemma \cite{HsiehDevetakWinter:08p} (see Ref. \onlinecite[Lem. 3]{Pereg:19a3}), %
there exists a POVM $\Lambda^0_{\htm_0}$ such that
\begin{align}
\label{eq:QpackB}
  \trace\left( \Lambda^0_{m_0} \rho_{B_1^n}^{x_0^n(m_0),x_1^n(m_0,m_1)} \right)  \geq 1-2^{-n[ H(B_1)_\rho-H(B_1|X_0 X_1)_\rho-(R_0+R_1)-\eps_3(\delta)]}
\end{align}
for all %
$m_0\in [1:2^{nR_0}]$. Hence,
$%
\cprob{ \mathscr{E}_2 }{ \mathscr{E}_1^c } \leq 2^{ -n( I(X_0,X_1;B_1)_\rho -(R_0+R_1)-\eps_3(\delta)) } 
$, %
which tends to zero as $n\rightarrow\infty$, provided that 
\begin{align}
R_0+R_1< I(X_0,X_1;B_1)_\rho -\eps_3(\delta) \,.
\label{eq:B2a}
\end{align}

Moving to the third term in the RHS of (\ref{eq:PeBsc}), suppose that $\mathscr{E}_2^c$ occurred, namely the decoder measured the correct $M_0$. Denote the state of the systems $B_1^n$ after this measurement by $\rho'_{B_1^n}$.
Then, as in Refs. \onlinecite{Pereg:19a3,Pereg:20c1}, we observe that due to the packing lemma inequality (\ref{eq:QpackB}),  %
the gentle measurement lemma \cite{Winter:99p,OgawaNagaoka:07p} implies that the post-measurement state is close to the original state in the sense that
\begin{align}
\frac{1}{2}\norm{\rho'_{B_1^n}-\rho_{B_1^n}}_1 \leq 2^{ -n\frac{1}{2}( I(X_0,X_1;B_1)_\rho -(R_0+R_1)-\eps_4(\delta)) } \leq \eps_5(\delta)
\end{align}
for sufficiently large $n$ and rates as in (\ref{eq:B2a}).
Therefore, the distribution of measurement outcomes when $\rho'_{B_1^n}$ is measured is roughly the same as if the POVM 
$\Lambda^0_{\htm_0}$ was never performed. To be precise, the difference between the probability of a measurement outcome $\htm_1$ when $\rho'_{B_1^n}$ is measured and the probability when $\rho_{B_1^n}$ is measured is bounded by $ \eps_5(\delta)$ in absolute value (see Ref. \onlinecite[Lem. 9.11]{Wilde:17b}).
Furthermore, 
\begin{align}
\trace\left[ \Pi^{\delta}(\rho_{B_1}|x_0^n,x_1^n) \rho_{{B_1}^n}^{x_0^n,x_1^n} \right] \geq& 1-\eps_6(\delta) \\
\Pi^{\delta}(\rho_B|x_0^n)  \rho_{{B_1}^n}^{x_0^n,x_1^n}   \Pi^{\delta}(\rho_{B_1}|x_0^n) \preceq& 2^{ -n(H({B_1}|X_0)_{\rho}-\eps_6(\delta)) } \Pi^{\delta}(\rho_{B_1}|x_0^n)
\\
\trace\left[ \Pi^{\delta}(\rho_{B_1}|x_0^n,x_1^n)  \right] \leq& 2^{ n(H({B_1}|X_0,X_1)_{\rho} +\eps_6(\delta))} \\
\trace\left[ \Pi^{\delta}(\rho_{B_1}|x_0^n) \rho_{{B_1}^n}^{x_0^n,x_1^n} \right] \geq& 1-\eps_6(\delta) 
\end{align}
for all $(x_0^n,x_1^n)\in\Aset^{\nicefrac{\delta}{2}}(p_{X_0} p_{X_1|X_0})$, by (\ref{eq:UnitTCond}), (\ref{eq:rhonProjIneqCond}), (\ref{eq:PidimCond}), and 
(\ref{eq:UnitTCondB}), respectively. Therefore, we have by the quantum packing lemma that there exists a POVM $\Lambda^1_{\htm_{1}|m_0}$ such that
$%
\cprob{ \mathscr{E}_3 }{ \mathscr{E}_1^c\cap  \mathscr{E}_2^c } \leq 2^{ -n( I(X_1;B_1|X_0)_\rho -R_1-\eps_7(\delta)) } 
$, %
which tends to zero as $n\rightarrow\infty$, provided that
\begin{align}
R_1< I(X_1;B_1|X_0)_\rho -\eps_7(\delta) \,.
\label{eq:B3}
\end{align}

It remains to consider erroneous decoding by Bob 2. Suppose that $\mathscr{E}_3^c$ occurred, namely Bob 1 measured the correct $m_{0}$, and thus sent the correct bin index $G$ such that $m_0\in\mathscr{B}(G)$. Recall that the size of each bin is $|\mathscr{B}(G)|=2^{n(R_0-\inC_{12})}$ (see (\ref{ref:bindef})). 
 Then, observe that
\begin{align}
\Pi^{\delta}(\rho_{B_2})  \rho_{B_2^n}   \Pi^{\delta}(\rho_{B_2}) \preceq& 2^{ -n(H(B_2)_{\rho}-\eps_2(\delta)) } \Pi^{\delta}(\rho_{B_1})
\\
\trace\left[ \Pi^{\delta}(\rho_{B_2}|x_0^n) \sigma_{{B_2}^n}^{x_0^n} \right] \geq& 1-\eps_2(\delta) \\
\trace\left[ \Pi^{\delta}(\rho_{B_2}|x_0^n)  \right] \leq& 2^{ n(H({B_2}|X_0)_{\rho} +\eps_2(\delta))} \\
\trace\left[ \Pi^{\delta}(\rho_{B_2}) \rho_{{B_2}^n}^{x_0^n} \right] \geq& 1-\eps_2(\delta) 
\end{align}
for all $x_0^n\in\Aset^{\delta}(p_{X_0,X_1})$, by (\ref{eq:rhonProjIneq}), (\ref{eq:UnitTCond}),  (\ref{eq:PidimCond}), and 
(\ref{eq:UnitTCondB}), respectively. Hence, by the quantum packing lemma (Ref. \onlinecite[Lem. 3]{Pereg:19a3}), %
there exists a POVM $\Gamma_{\tm_0}$ such that
$%
\cprob{ \mathscr{E}_4 }{ \mathscr{E}_1^c\cap \mathscr{E}_2^c\cap \mathscr{E}_3^c } \leq 2^{ -n( I(X_0;B_2)_\rho -(R_0-\inC_{12})-\eps_8(\delta)) } 
$, %
which tends to zero as $n\rightarrow\infty$, provided that 
\begin{align}
R_0< I(X_0;B_2)_\rho+\inC_{12} -\eps_8(\delta) \,.
\label{eq:B2}
\end{align}
To show that rate pairs in $\frac{1}{\kappa}\mathcal{R}_{\text{Cl}}(\channel^{\otimes \kappa})$ are achievable as well, one may employ the coding scheme above for the product broadcast channel $\channel^{\otimes \kappa}$, where $\kappa$ is arbitrarily large.
This completes the proof of the direct part.

\subsection{Converse Proof}
Consider the converse part for the regularized capacity formula. %
 Suppose that Alice chooses $m_0$ and $m_1$ uniformly at random, and prepares an input state $\rho^{m_0,m_1}_{ A^n}$. 
After Alice sends the systems $A^n$ through the channel,  the output state is
$\rho_{B_1^n B_2^n}=\frac{1}{2^{n(R_0+R_1)}}\sum_{m_0,m_1}^{2^{nR_0}} \sum_{m_1=1}^{2^{nR_1}} \channel_{A^n\rightarrow B_1^n B_2^n}(\rho^{m_0,m_1}_{ A^n}) $.
Then, Bob 1 performs a decoding POVM $\Lambda^{m_0,m_1,g}_{B_1^n}$, sends $g$ to Bob 2 using conferencing, and Bob 2 chooses a POVM
$\Gamma^{m_0}_{B_2^n|g}$ accordingly.
Consider a sequence of codes $(\Fset_n,\Lambda_n,\Gamma_n)$ such that the average probability of error tends to zero, hence
the error probabilities $\prob{ \hM_0\neq M_0 }$, $\prob{ (\hM_0,\hM_1)\neq (M_0,M_1)}$, $\prob{ \hM_1\neq M_1 |M_0}$ are bounded by some
$\alpha_n$ which tends to zero as $n\rightarrow \infty$.
By Fano's inequality \cite{CoverThomas:06b}, it follows that%
\begin{align}
H(M_0|\tM_0) \leq n\eps_n\\
H(M_0,M_1|\hM_0,\hM_1) \leq n\eps_n'\\
H(M_1|\hM_1,M_0) \leq n\eps_n''
\end{align}
where $\eps_n,\eps_n',\eps_n''$ tend to zero as $n\rightarrow\infty$.
Hence, 
\begin{align}
nR_0&= H(M_0)=I(M_0;\tM_0)_{\rho}+H(M_0|\tM_0) \leq I(M_0;\tM_0)_{\rho}+n\eps_n \nonumber\\
&\leq I(M_0;B_2^n G)_{\rho}+n\eps_n= I(M_0;B_2^n)_{\rho}+I(M_0;G|B_2^n)_\rho+n\eps_n \nonumber\\
&\leq I(M_0;B_2^n)_{\rho}+n\inC_{12}+n\eps_n
\label{eq:ConvIneq1SC}
\intertext{where the second inequality %
follows from the Holevo bound (see Ref. \onlinecite[Theo. 12.1]{NielsenChuang:02b}), and the last inequality holds as $I(M_0;G|B_2^n)_\rho\leq H(G)\leq n\inC_{12}$ because $G$ is a \emph{classical} message in $[1:2^{n\inC_{12}}]$. Similarly,}
n(R_0+R_1)&\leq I(M_0,M_1;B_1^n)_{\rho}+n\eps_n' .
\label{eq:ConvIneq2SC}
\end{align}

Furthermore, since $M_0$ and $M_1$ are statistically independent, we can also write 
\begin{align}
nR_1= H(M_1|M_0) \leq& I(M_1;\hM_1|M_0)_{\rho}+n\eps_n'' %
\leq I(M_1;B_1^n|M_0)_{\rho}+n\eps_n'' \,.
\end{align}
We deduce that $R_0\leq I(X_0^n;B_2^n)_{\rho}+\inC_{12}+\eps_n $, $R_0+R_1\leq I(X_0^n,X_1^n;B_1^n)_{\rho}+\eps_n' $, and
$R_1\leq I(X_1^n;B_1^n|X_0^n)_{\rho}+\eps_n'' $
with $X_k^n=f_k(M_k)$ where $f_k$ are arbitrary one-to-one maps from $[1:2^{nR_k}]$ to $\Xset_k^n$,
for $k=0,1$. 
This completes the converse proof for the regualarized characterization.

For a Hadamard broadcast channel, where Bob 1 receives a classical output $Y_1^n$, define
\begin{align}
X_{0,i}=(M_0,Y_1^{i-1}) \,,\; X_{1,i}=(M_1,Y_1^{i-1}) \,.
\end{align}
Applying the chain rule to (\ref{eq:ConvIneq1SC}),
\begin{align}
nR_0 &\leq %
\sum_{i=1}^n I(M_0;B_{2,i}|B_2^{i-1})_{\rho}+n\inC_{12}+n\eps_n \leq \sum_{i=1}^n I(M_0 B_2^{i-1} ;B_{2,i})_{\rho}+n\inC_{12}+n\eps_n. 
\end{align}
Since the marginal of Bob 2 is degraded with respect to that of Bob 1, namely $\channel^{\,\text{H}}_{A\rightarrow B_2}=\Pset_{Y_1\rightarrow B_2}\circ\channel^{\,\text{H}}_{A\rightarrow Y_1}$, the data processing inequality for the quantum mutual information implies that $I(M_0 B_2^{i-1} ;B_{2,i})_{\rho}\leq I(M_0 Y_1^{i-1} ;B_{2,i})_{\rho}=I(X_{0,i} ;B_{2,i})_{\rho}$, hence
\begin{align}
R_0-\eps_n &\leq %
\frac{1}{n}\sum_{i=1}^n I(X_{0,i};B_{2,i})_{\rho}+\inC_{12}= I(X_{0,K};B_{2,K}|K)_{\rho}+\inC_{12}
\end{align}
where  $K$ is a classical random variable with uniform distribution over $[1:n]$, independent of $M_0$, $M_1$, and $G$. Defining
$\rho_{KA}=\frac{1}{n}\sum_{i=1}^n \kb{i}\otimes \rho_{A_i}$, $\rho_{KB_1 B_2}=\channel_{A\rightarrow B_1 B_2}(\rho_{KA)}$, and
\begin{align}
X_0\equiv (X_{0,K},K) \,,\; X_1\equiv (X_{1,K},K)
\end{align} 
we obtain 
\begin{align}
R_0 &\leq I(X_{0};B_{2})_{\rho}+\inC_{12}+\eps_n 
\intertext{and by similar considerations,}
R_0+R_1 &\leq I(X_0 X_1;B_1)+\eps_n \\
R_1&\leq I(X_1;B_1|X_0)+\eps_n  %
\end{align}
This completes the proof of Theorem~\ref{theo:BCcl}.
\qed %

\bibliography{references2}%

\end{document}